\documentclass[prd,twocolumn,amsmath,nobibnotes,nofootinbib,superscriptaddress]{revtex4}

\usepackage{bm} \usepackage{graphicx}

\def\ba{\begin{eqnarray}}
\def\ea{\end{eqnarray}}
\def\be{\begin{equation}}
\def\ee{\end{equation}}
\def\({\left(}
\def\){\right)}
\def\[{\left[}
\def\]{\right]}
\def\<{\left<}
\def\>{\right>}

\begin{document}

\title{Towards observable signatures of other bubble universes II: Exact solutions for thin-wall bubble collisions}
\date{\today}

\author{Anthony Aguirre}
\email{aguirre@scipp.ucsc.edu}
\affiliation{SCIPP, University of California, Santa Cruz, CA 95064, USA}
\author{Matthew C Johnson}
\email{mjohnson@theory.caltech.edu}
\affiliation{California Institute of Technology, Pasadena, CA 91125, USA}

\begin{abstract}
We assess the effects of a collision between two vacuum bubbles in the thin-wall limit. After describing the outcome of a generic collision possessing the expected hyperbolic symmetry, we focus on collisions experienced by a bubble containing positive vacuum energy, which could in principle contain our observable universe. We provide criteria governing whether the post-collision domain wall accelerates towards or away from this ``observation" bubble, and discuss the implications for observers located at various positions inside of the bubble. Then, we identify the class of solutions which have minimal impact on the interior of the observation bubble, and derive a simple formula for the energy density of a shell of radiation emitted from such a collision. In the context of a universe undergoing false vacuum eternal inflation, these solutions are perhaps the most promising candidates for collisions that could exist within our past light cone, and therefore in principle be observable.
\end{abstract}

\maketitle

\section{Introduction}

Eternal inflation, once widely regarded as an extravagant curiosity (except by most of inflation's pioneers!) has become increasingly central in cosmology. Observationally, evidence for both present-time accelerated cosmic expansion and for early inflation suggest (at least) two epochs of vacuum-energy domination.  Theoretically, many theories of particle physics beyond the standard model seeking to account for these vacuum energies lead to metastable vacua in the potential energy of a scalar field. This feature (as well as local maxima or nearly-flat regions of the potential) can drive eternal inflation, by which local regions with potentially diverse properties (``pocket universes") are produced from a de Sitter (dS)-like background inflating spacetime that admits a foliation in which inflation is future-eternal. (For recent reviews, see, e.g.,~\cite{Linde:2007fr,Aguirre:2007gy,Guth:2007ng,Vilenkin:2006ac}).

Because an (undisturbed) pocket universe is separated from the inflating background by an infinite spacelike reheating surface, it is generally assumed that other supposed pocket universes could have no observational impact on ours. Coupled with the tendency of inflation to erase information about prior epochs, this has led to a widespread belief that eternal inflation has no directly\footnote{If the observed properties of our universe are determined by the statistical distribution of properties across the multiverse along with selection effects or other required conditions, there may be hints of this fact in the values of, or correlations between, those observables; see e.g.\cite{Aguirre:2004qb,Aguirre:2005cj} for discussion.} observable signatures.

This, however, may be overly pessimistic.  Let us consider models of ``false vacuum" eternal inflation (FVEI), driven by sufficiently long-lived metastable vacua. In these models, tunneling out of the metastable minimum is assumed to proceed via the Coleman-de Luccia (CDL) instanton~\cite{Coleman:1980aw}\footnote{Depending on the details of the potential landscape, the transition can be mediated by a number of other mechanisms~ \cite{Hawking:1981fz,Farhi:1989yr,Garriga:2004nm}, but if the CDL instanton exists, it is the most probable. Whether the existence of the CDL instanton is common is a difficult issue requiring a measure over both inflationary transitions~\cite{Aguirre:2006na} and over potential shapes.}. The spacetime described by the Lorentzian instanton possesses O(3,1) invariance and contains regions identifiable with an infinite open Friedmann-Lemaitre-Robertson-Walker (FLRW) universe~\cite{Coleman:1980aw}. Such models, which go by the name of ``open inflation", have been well-explored in the literature (e.g.,~\cite{Gott:1982zf,Bucher:1994gb,Linde:1995rv,Yamamoto:1995sw,Linde:1998iw,Garriga:1998he,Linde:1999wv}), and with an appropriate (though not necessarily ``natural") choice of scalar potential can be completely consistent with the variety of precision cosmological observables to which we presently have access.

What observational signatures might such models have? If we assume a short epoch of inflation within the bubble (for which there may be theoretical bias~\cite{Freivogel:2005vv}), then open inflation has several signatures observable in the CMB, including negative cosmological curvature and wall fluctuations that propagate into the bubble (see, e.g.,~\cite{Garriga:1997ht,Garriga:1998he}). However, even if open inflation could {\em only} arise via decay from a false vacuum (and this is questionable: see~\cite{1998PhLB..425...25H}), this would only provide circumstantial evidence for eternal inflation and other pocket universes.

In an effort to do better, inspired by the work of Ref.~\cite{Garriga:2006hw}, we discussed in a previous publication~\cite{Aguirre:2007an} (hereafter AJS) a possible {\em direct} signature of eternal inflation: the effect of collisions between bubble universes. This study computed the probability and angular size of bubble collisions in an observer's past lightcone, in the limit where a collision has no effect on the interior of the ``observation bubble" (which would contain our observable universe).  While observable collisions would be generic for nucleation rates $\Lambda$ (per unit 4-volume) of order $H_F^{-4}$ (barely satisfying the criterion for eternal inflation), in the more expected case of exponentially suppressed rates, the study indicated that there are two classes of collision events that might nonetheless be seen with high probability.  ``Late time" events (which typically enter the observer's past light cone at cosmological times $\tau \gg H_F^{-1}$) would be isotropically distributed, and would affect a small or negligible solid angle on the sky. (However, the total number of such events depends much more sensitively on the assumed cosmology inside of the bubble than assumed in AJS, a point which we clear up in Appendix~\ref{bubblecosmo}). The second class, ``early time" collisions, are strongly anisotropic (in agreement with the analysis of~\cite{Garriga:2006hw}), cover the entire sky, and and are seen by essentially all observers independent of the cosmology inside of the bubble.

While encouraging for the prospect of observing other bubble universes, this study did not address the detailed effects of collisions on the interior of the observation bubble, so key questions remain open. To what extent are the assumed symmetries of the open FLRW universe inside of the bubble preserved in the presence of collisions? Could we live to the future of a collision event, and could its effects be observed not just in principle but in practice? What would those effects look like in the CMB, 21cm radiation, or other observables?  In this paper, as a step towards answering these questions, we analyze exact solutions forming a simplified model of the collision between two bubbles.\footnote{Near the end of this work's preparation, the manuscript~\cite{Chang:2007eq} appeared, which also solves essentially the same thin-wall bubble collision problem of Sec.~\ref{postcollision} below.}

From the SO(3,1) symmetry of a single bubble,\footnote{The instanton describing bubble nucleation has O(4) symmetry and can be continued into an O(3,1)-symmetric Lorentzian instanton describing a time-symmetric bubble. An actual bubble nucleated within a background spacetime will not have this symmetry (as the matching surface breaks it), but the region of interest, after the nucleation and largely within the event's future lightcone, can be considered as SO(3,1)-symmetric.} an SO(2,1) symmetry remains in the collision of two bubbles.  This places strong constraints on the post-collision spacetime via a version of Birkhoff's theorem that allows one to write down the set of all metrics with this symmetry.  Following previous work~\cite{Freivogel:2007fx,Wu:1984eda,Blanco-Pillado:2003hq,Berezin:2002nc,Bousso:2006ge,BlancoPillado:2001si,Hawking:1982ga,Bucher:2001it,Moss:1994pi}, we model the post-collision spacetime by considering thin-wall junctions between such metrics. If the phase in each of the colliding bubbles is different, a domain wall must form separating the bubble interiors. Energy and momentum conservation at the collision dictate that there must be other energy sinks as well, which we model as shells of radiation emanating from the collision event. If the phases are the same, then radiation must be emitted from the collision, but no domain wall need form.

As emphasized in AJS, if there are {\em any} collision types for which the post-collision bubble interior is compatible with standard cosmological evolution and admits infinite spacelike slices of nearly-homogeneous density, then independent of nucleation rate, {\em all but a measure zero of observers will have an ``early time" collision to their past}. It is therefore important to identify those collision types that will minimally disturb the interior of the observation bubble, taking into account the intrusion of a post-collision domain wall, the backreaction of the collision on the spacetime, and the intensity of the emitted radiation. Thus a major goal of this study is to explicitly construct solutions for the model collisions that can be regarded as the most optimistic candidates for observable collisions.

In Sec.~\ref{frames}, we introduce three reference frames in which it will be important to understand the effects of bubble collisions and their relation to the standard picture of eternal inflation. We set up the collision between two thin-wall bubbles with arbitrary characteristics (vacuum energy and tension) in a background de Sitter space in Sec.~\ref{modelproblem}, and then discuss the formalism necessary to determine the form of the post-collision spacetime in Sec.~\ref{postcollision}. Finally, in Sec.~\ref{detailedcoll}, we identify the most promising candidates for observable bubble collisions and derive a simple formula for determining the energy density of radiation emitted into the observation bubble and conclude in Sec.~\ref{conclusions}. We also include three appendices: Appendix~\ref{hyperspaces} describes the general form of hyperbolic vacuum spacetimes with a cosmological constant, Appendix~\ref{sec-appptol} in which we analyze the frame-dependence of the distribution of the dS-invariant separation between colliding bubbles, and Appendix~\ref{bubblecosmo} where we analyze the expected number of collisions for an arbitrary cosmology inside of the bubble. We work in natural units unless otherwise noted.

\section{Eternal inflation, collisions, and symmetries}\label{frames}

\subsection{Open inflation and false-vacuum eternal inflation}

For detailed accounts of open inflationary bubbles and the issue of collisions between them, we refer the reader to AJS, which is very useful prior reading for this paper; for a general and slightly more pedagogical recent account of eternal inflation see~\cite{Aguirre:2007gy}.

Very briefly, in an inflation potential $V(\phi)$ with false (higher) and true (lower) vacua, the false vacuum drives exponential expansion leading to a dS-like background spacetime described by metric
\begin{equation}\label{eq-flatds}
ds^2 = -d t^2 +e^{2H_F t} \left[d r^2 + r^2 \ d\Omega_{2}^{2} \right], 
\end{equation}
where $d\Omega^2=d\theta^2+\sin^2\theta d\phi^2$, and where $0 \le r < \infty$, $-\infty < t < \infty$, $0 \leq \theta \leq \pi$, and $0 \leq \phi \leq 2 \pi$.
Within this background, if the potential is of suitable form, bubbles nucleate as per the CDL instanton, so that within a null cone emanating to the future of the nucleation event, there is an open FLRW region with metric
\begin{equation}\label{opends}
ds^2 = -d \tau^2 +a^2(\tau) \left[d \xi^2 + \sinh^2 \xi \ d\Omega_{2}^{2} \right],
\end{equation}
where equal-$\tau$ surfaces are homogeneous spacelike hyperboloids and coincide with surfaces of constant $\phi$. As $\tau$ increases, the field $\phi$ evolves from the tunneled-to value, through a (presumed) inflation phase, until reheating and subsequent standard big-bang cosmological evolution. The background spacetime is never completely converted into bubbles; rather, the remainder assumes instead a steady-state form that is a fractal of dimension $<3$~\cite{Vilenkin:1992uf}.  

\subsection{Bubble collisions in flat space}
In order to gain some intuition, let us analyze the collision of two bubbles in a background Minkowski space. Imagine that the nucleation centers of two identical bubbles are positioned at $(x= \pm b , y=0, w=0, t=0)$. By O(3,1) symmetry, the wall equation of motion for each bubble is then
\begin{equation}\label{eq-minkbub}
(x \pm b)^2 + y^2 + w^2 -t^2 = R_0^2
\end{equation}
where $R_0$ is the initial radius of the bubble. As the bubbles grow, they meet at $x=0$, and the collision surface traces out a hyperboloid $y^2 + w^2 - t^2 = (R_0^2 - b^2)$ in the $(x=0,y,w,t)$ plane centered around $(y=0,w=0)$. If the bubbles form at different times, or meet at a position other than $x=0$, the collision still traces out a hyperboloid in the $y-w$ plane, because a boost and a translation can always bring us to the frame described by Eq.~\ref{eq-minkbub}. Therefore, it makes sense to take advantage of the hyperbolic symmetry in the plane of the collision. The coordinate transformation
\begin{eqnarray}
&& t = z \cosh \chi,\ \  x = x, \\ \nonumber
&& y = z \sinh \chi \cos \phi, \ \  w = z \sinh \chi \sin \phi
\end{eqnarray}
makes the hyperbolic symmetry of the metric manifest:
\begin{equation}
\label{eq-minkhyp}
ds^2 = -dz^2 + dx^2 + z^2 dH_{2}^{2},
\end{equation}
where $
dH_{2}^{2} = d \chi^2 + \sinh^2 \chi d\phi^2$ is the metric of a spacelike 2-hyperboloid.

\begin{figure}
\includegraphics[width=7.5cm]{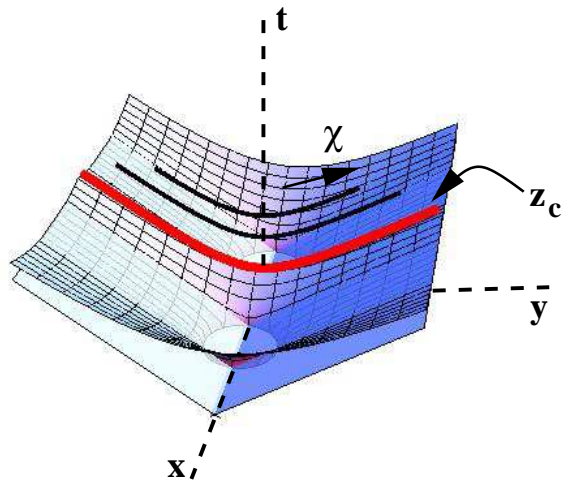}
\caption{The collision of two bubbles in Minkowski space with nucleation centers located on the $x-$axis. The hyperbolic coordinates Eq.~\ref{eq-minkhyp} cover the portion of the spacetime above the planes, and the collision surface, outlined by the thick red line, is located at a constant hyperbolic position $z_c$. Moving along the constant $z$ hyperboloids corresponds to increasing $\chi$ in Eq.~\ref{eq-minkhyp}.
  \label{fig-minkcoll}
}
\end{figure}

In this coordinate system, the intersection between the colliding walls occurs at $z= {\rm const}$. The coordinate patch does not cover the entirety of Minkowski space, but only $w^2 + y^2 > t^2$ for all $x$ ($-\infty < x < \infty $), as illustrated in Fig.~\ref{fig-minkcoll}.\footnote{However, we can continue across $z=0$ to a system where the hyperboloids are timelike, and cover the entirety of Minkowski space patch-wise.} If we neglect the backreaction of the fields inside of the bubbles on the geometry, then this patch covers a portion of the bubble interiors as well, fortunately including the entire causal future of the collision region. This simple example illustrates that there are important symmetries we can take advantage of in the collision spacetime: the task of determining the effects of bubble collisions is a two-dimensional problem involving $x$ and $z$. Every further example we will study has this symmetry, and we will be able to generalize the simple picture of Fig.~\ref{fig-minkcoll} to collisions occurring in and giving rise to more complicated spacetimes.

\subsection{Bubbles and their collisions in dS}

As in Minkowski space, we can choose coordinates for dS adapted to the symmetries of the problem, in which the metric takes the form:
\begin{eqnarray}
\label{eq-dshyp}
ds^2 =  - (1+H^2 z^2)^{-1} dz^2 + (1+H^2 z^2) dx^2 + z^2 dH_{2}^{2} 
\end{eqnarray}
with $0 \leq z < \infty$, $0 \leq H x \leq \pi$, $0 \leq \chi < \infty$, and $0 \leq \phi \leq 2 \pi$, which manifestly approaches the Minkowski form of Eq.~\ref{eq-minkhyp} as $H\rightarrow 0$.  Treating a collision between bubbles in a fixed dS background would, using this metric, be exactly analogous to the Minkowski case.

However, we wish to treat exact solutions that join bubble interiors to each other and to the background dS via domain walls.  In doing so it is very instructive to take an embedding-space picture.
First, consider the background dS space out of which the colliding bubbles nucleate. The entire manifold is represented by the surface of a hyperboloid ${\cal H}$ defined by $\eta_{\mu\nu}X^\mu X^\nu =H^{-2}$ in the embedding Minkowski space with coordinates $X^\mu$ ($\mu=0..4$). Here, the O(4,1) symmetry of dS is manifested as the Lorentz group in 5-D.  

There are several useful coordinate systems on dS.  Of greatest use here will be the ``flat foliation" with metric given by Eq.~\ref{eq-flatds},  and the ``closed foliation" with metric
\begin{equation}\label{closedds}
ds^2 = \frac{1}{H^{2} \cos^2 T} \left[-dT^2 + d\eta^2 + \sin^2 \eta \ d\Omega_{2}^{2} \right]
\end{equation}
(where $-\pi / 2 \leq T \leq \pi / 2$, $0 \leq \eta \leq \pi$).  The latter metric is induced by coordinatizing the hyperboloid as:
\begin{eqnarray}
\label{eq-closedembedding}
&& X_{0} = H^{-1} \tan T \\ \nonumber
&& X_{i} = H^{-1} \frac{\sin\eta}{\cos T} \omega_{i},\ \ X_{4} = H^{-1} \frac{\cos \eta}{\cos T},
\end{eqnarray}
with $(\omega_1,\omega_2,\omega_3)=(\cos\theta,\sin\theta\cos\phi,\sin\theta\sin\phi)$.

To include a single bubble separated from this background by a thin domain wall, we can define the wall via the intersection of ${\cal H}$ with an appropriate timelike plane.  (Note that this explicitly breaks O(4,1) down to O(3,1), the expected symmetry of the one-bubble spacetime.) For example, for a bubble nucleated about one pole at the ``throat" of dS ($T=0$ and $\eta=0$ in the closed foliation), this plane is at constant $X_4$, with $0 \le X_4 \le H_F^{-1}$. Within the bubble lies a spacetime with metric
\begin{equation}\label{eq-openfrw}
ds^2 = \mp d \tau^2 +a^2(\tau) (dH_{3}^2)^{S,T}
\end{equation}
This can be induced via the embedding
\begin{eqnarray} \label{eq-openslices}
X_{0} &=& a(\tau) g_0^{S,T} (\xi),\\ \nonumber
 X_{i} &=& a(\tau) g_1^{S,T} (\xi)  \omega_{i},\ \ X_{4} = f (\tau),
\end{eqnarray}
where $g_{0}^{S,T} (\xi) =  (\cosh \xi, \ \sinh \xi)$, $g_{1}^{S,T} (\xi) =(\sinh \xi, \ \cosh \xi)$, and $f(\tau)$ solves $f'^2(\tau)=a'^2(\tau)-1$. 
Different values of $X_4$ correspond to 3D hyperboloids of constant field value in the Lorentzian CDL instanton.  Outside the null cone $X_4=H^{-1}$ these are timelike (corresponding to the bottom signs and ``T" superscript), while inside the null cone (top signs and ``S" superscript) they are spacelike surfaces that correspond to equal-time slices in the FLRW cosmology. Note that if we set $a(\tau)= (H_T^{-1} \sinh H_T \tau, H_T^{-1} \cosh H_T \tau)$, we recover empty dS of radius $H_T^{-1}$ in the ``open foliation."

Now let us contemplate two colliding bubbles.  In the embedding space, adding another bubble corresponds to cutting  ${\cal H}$ with an additional plane, across which a different embedded spacetime is matched. By cutting the hyperboloid again, we have further reduced the symmetry of the spacetime from O(3,1) to O(2,1). The triple intersection of the two planes and ${\cal H}$ represents the collision surface, and the embedding space picture in the future light cone of the collision will in general not be well defined (because one is not guaranteed an embedding in $n+1$ dimensions \cite{Kasner:1921fk}).

\section{The model problem}\label{modelproblem}

\begin{figure*}[htb]
\includegraphics[width=14 cm]{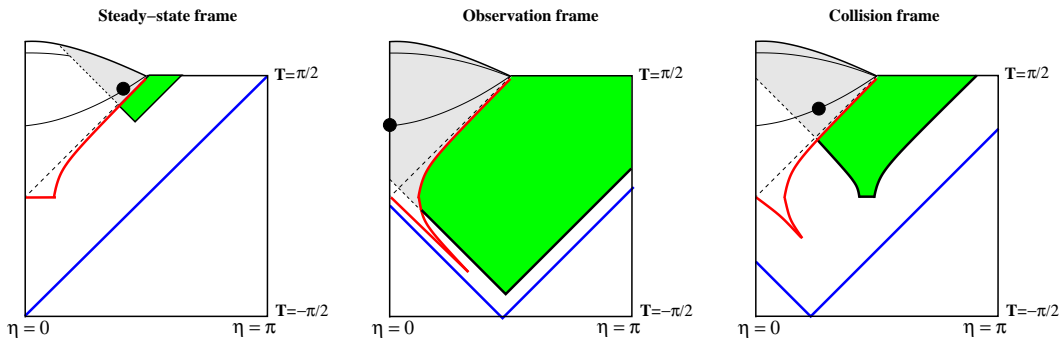}
\caption{The three frames described in Sec.~\ref{sec-refframes} as shown in a ``conformal slice" of constant $\theta,\phi$. The observation bubble is on the left, and the colliding bubble on the right. A hypothetical observer is denoted by the dot, and regions inside of the observation bubble to the future of the collision event are shaded grey (we will determine the structure of the post-collision spacetime in later sections). The initial value surface (along the $(\theta,\phi)$ direction separating the nucleation centers) in the background false vacuum is indicated by the solid blue line. In the limit where the interior of the observation bubble remains undisturbed, the transformation between frames is accomplished by boosting in the embedding space, which moves points in the background false vacuum as in Fig.~\ref{fig-dssymm}. This brings the colliding bubbles to earlier times, and stretches the observation bubble wall below $T=0$. 
  \label{fig-frames}
}
\end{figure*}

A number of previous works have studied bubble collisions; see, e.g.,~\cite{Chang:2007eq,Freivogel:2007fx,Wu:1984eda,Blanco-Pillado:2003hq,Berezin:2002nc,Bousso:2006ge,BlancoPillado:2001si,Hawking:1982ga,Bucher:2001it,Moss:1994pi}. Here we investigate in detail a simplified model in which both the bubbles and the collision region are pure vacuum, and all regions are joined by thin walls. First, we specify the vacuum solutions and their junctions.  Second, we discuss the embedding of these into the cosmological context, and the different frames in which it is useful to analyze the scenario. Then we specify initial conditions for the collision and enumerate the input parameters.  The solution to this model problem then occupies Sec.~\ref{postcollision} and~\ref{detailedcoll}.

\subsection{Exact solutions modeling the bubble collision}\label{exactsolutions}

The simplified problem that we shall treat is one in which:

\begin{enumerate}

\item The background space time is dS, with Hubble constant $H_F$.

\item The ``observation" bubble is (prior to the collision) either dS or Minkowski, with Hubble constant $H_o$, and nucleates with proper radius $R_o^i$.  It is joined to the background by a thin domain wall of tension $k_o$ (related to the energy-momentum tensor of the wall as per $k_o=4\pi G\sigma_o$ and Eq.~\ref{eq-ktuv} below).

\item The ``collision" bubble is dS, AdS or Minkowski, with Hubble constant $H_C$\footnote{Note that using our conventions, $H^2=0$ for Minkowski and $H^2 < 0$ for AdS.}, and nucleated with proper radius $R_C^i$, joined to the background by a wall of tension $k_C$.

\item The post-collision region is a vacuum region with SO(2,1) symmetry modeled as two vacuum regions joined by a wall of tension $k_{oC}$.  These regions have Hubble parameters $H_o$ and $H_C$ like the colliding bubbles, but may also have mass parameters (denoted $M_o$ and $M_C$) which appear in the hyperbolic vacuum solutions reviewed in Appendix~\ref{hyperspaces}.  As noted below (see also~\cite{Freivogel:2007fx}), to conserved energy-momentum in a general collision an extra ``energy sink" is required.  Inspired by numerical solutions \cite{Hawking:1982ga,Blanco-Pillado:2003hq} and physical considerations, we follow ~\cite{Freivogel:2007fx,Wu:1984eda,Blanco-Pillado:2003hq,Bousso:2006ge,BlancoPillado:2001si,Hawking:1982ga,Bucher:2001it} and model this as an outgoing null shell of radiation that joins the post-collision region to the pre-collision bubble interiors. The energy density of this shell will, as we shall see, be determined by the collision kinematics.

\end{enumerate}

\subsection{Reference frames}\label{sec-refframes}
It will be helpful to define three classes of observers and their associated reference frames.
\begin{itemize}
\item {\bf Steady-state frame} This is the frame defined by the steady state distribution of bubbles in the eternally inflating false vacuum spacetime (e.g.,~\cite{Vilenkin:1992uf,Aguirre:2001ks,Aguirre:2003ck}; note that this frame endures indefinitely after any boundary conditions surface as explicitly shown by~\cite{Garriga:2006hw}). The observation bubble nucleates at flat slicing $t=0$ and the initial conditions surface is located at flat slicing $t \rightarrow -\infty$ (see~\cite{Garriga:2006hw} and AJS).  This frame is most useful for connecting bubbles to the background eternally inflating spacetime.
\item {\bf Observation frame} This is the frame defined by our assumed position inside of the observation bubble. The observer is located at $x=0$ in this frame (which at large $z$ corresponds to the center of the bubble by convention), but the initial value surface is distorted due to a relative boost between this frame and the steady-state frame. This frame is most useful for assessing what a given observer would experience.
\item {\bf Collision frame} In this frame, both the observation bubble and a colliding bubble are nucleated at global slicing $T=0$. The initial value surface is distorted, and the observer will generally not be located at the origin of the observation bubble.  This frame is most useful for computing the results of a bubble collision event.
\end{itemize} 

For simplicity, and to set up the collision problem we will solve later, we restrict this description to spacetimes with only two bubbles as shown in Fig.~\ref{fig-frames}, although these frames are more generally well-defined. In each cell of Fig.~\ref{fig-frames}, the position of the observer, the initial value surface, and the position of the colliding bubble are shown. The shaded region inside of the observation bubble to the future light cone of the collision indicates our present ignorance of the post-collision environment (which we will remedy in later sections). 

Outside of the collision region, transformations between the steady-state and collision frames can be defined entirely in terms of the background dS, independent of any detailed knowledge of how a collision will affect the bubble interiors. Specifying the observation frame is somewhat more complicated, and requires a detailed knowledge of the bubble interior. In order to define the observation frame, there must exist timelike trajectories inside of the observation bubble that remain for all times in the observation bubble (i.e., if the post-collision domain wall enters the observation bubble and cuts off timelike infinity, the observation frame is ill-defined). In all three cases, specifying what occurs in the collision region under a ``boost" transformation (that corresponds to a translation along an equal-time surface in an unperturbed bubble) is difficult and we leave this for future work. However, under the (unrealistic) assumption that the observation bubble is unaffected by the collision, the transformation between all three frames can be performed explicitly using Lorentz transformations in the embedding space; this will provide a qualitative understanding of how the three frames are related. 

\begin{figure}[htb]
\includegraphics[width=5cm]{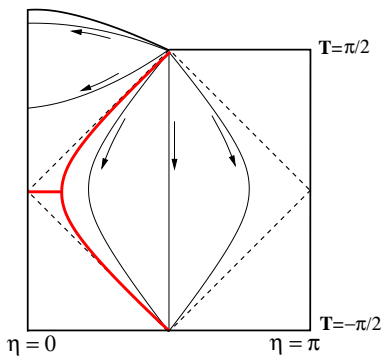}
\caption{The action of the boost Eq.~\ref{eq-embeddingboost} on points in the spacetime, with arrows pointing along the flow of increasing boost parameter $v$. A conformal slice with $\theta, \phi$ fixed in the direction joining the bubble nucleations is shown. The solid red line denotes the time-symmetric observation bubble wall. A boost applied to a bubble nucleated at $\eta=T=0$ (which is not time-symmetric) pushes the wall along the time-symmetric trajectory.
  \label{fig-dssymm}
}
\end{figure}

Let us consider this explicit transformation in a specific example where we choose a given direction for the incoming bubble. In each of the frames, the observation bubble wall is at a constant position $X_4=X_{4}^{\rm wall}$ (independent of $X_0$). Boosting the embedding space in directions parallel to this plane does not affect the position of the wall, but does non-trivially affect points inside and outside the bubble. We fix the wall of the colliding bubble to be defined by the intersection of the embedding space hyperboloid with an appropriate timelike plane (depending upon the frame and the junction conditions) specified by coordinates $(X_0,X_1,X_4)$.  Placing our observer along the direction separating the two nucleation centers,\footnote{For typical observers at large open slicing radii $\xi_o$, this is the direction from which we expect most observable collisions to originate.} we can transform between the various frames by the boost:
\begin{eqnarray}
\label{eq-embeddingboost}
&& X_{0}' = \gamma \left(X_{0} - v X_{1} \right), \\ \nonumber
&& X_{1}' = \gamma \left(X_{1} - v X_{0} \right), \\ \nonumber
&& X_{2,3,4}' = X_{2,3,4}.
\end{eqnarray}
where by convention, we start in the steady-state frame (unprimed coordinates),  and different boost parameters $v$ (with  $\gamma=(1-v^2)^{-1/2}$) are necessary to go to the collision and observation frames. In the latter case, the boost parameters $\gamma_o = \cosh \xi_{o}$ and $v_o = \tanh \xi_{o}$ translate an observer at $\xi_o$ to the origin of the observation bubble, while shifting the position of the colliding bubble. In the former, we must transform the intersecting plane such that it is independent of $X_0$ (vertical in the embedding space) or, equivalently, must take the global slicing time of the nucleation center of the colliding bubble to zero. The boost parameters are given in terms of the global-slicing position $(\eta_n, T_n)$ of the colliding bubble's nucleation center by $\gamma_C = \frac{\sin \eta_n}{\sqrt{\sin^2 \eta_n - \sin^2 T_n}}$ and $v_C = \frac{\sin T_n}{\sin \eta_n}$. The observer in this frame is generally located at an open slicing position  $\xi_o ' \neq 0$. Shown in Fig.~\ref{fig-dssymm} are the orbits of the boost in the background de Sitter space and inside of the observation bubble. A positive boost parameter $v$ will move points in the indicated directions along these orbits.

In Sec.~\ref{detailedcoll} we will return to the problem of transforming to and from the observation frame when the interior of the observation bubble is affected by collisions, but for the analysis we now present, we will focus on the description of two-bubble collisions in the collision frame. 
 
 \subsection{Before the collision}\label{beforecollision}

In the collision frame, the kinematical parameters necessary to specify the initial conditions for a bubble collision are the trajectories of the individual walls and the distance between the nucleation centers. We will choose a coordinate system in which the nucleation center of the observation bubble is located at $\eta = T = 0$ and the nucleation center of the colliding bubble is located at $(T=0, \eta = \eta_C, \theta=0,\phi=0 )$. The trajectory of the observation bubble wall can be found by using the condition that $X_{4}$ is a constant along the wall, with the exact position determined by solving the Israel junction conditions (e.g.,~\cite{Garriga:1997ef}) to give
\begin{equation}
X_{4} = H_o^{-1} (1 - H_o^2 {R^{i}_{o}}^{2})^{1/2},
\end{equation}
with 
\begin{equation}
\label{eq-initialrad}
R^{i}_{o} = \frac{2 k_o}{\left[ (H_F^2 + H_o^2 + k_o^2 )^2 - 4 H_F^2 H_o^2 \right]^{1/2}}.
\end{equation}
We can see how the initial radius is related to the scalar potential responsible for the CDL instanton as follows. Consider a potential of the form $V(\phi) = \mu^4 v(\phi / M)$, where we assume that there are only two relevant scales in the potential: $\mu$ characterizing the energy scale of the minima and barriers and $M$ characterizing the width of the potential barrier, with $v(\phi / M)$ consequently a function with amplitude of order unity and slowly varying in $\phi/M$. The tension of the bubble wall will be of order $\sigma \sim \mu^2 M$ and in the limit where the initial radius is small (and therefore gravitational effects are small)
\begin{equation}
R_o \sim \frac{M}{\mu^2 \left[ v(\phi_F / M) - v(\phi_T / M) \right]}
\end{equation}
where the denominator is the dimensionless energy splitting between the true and false vacuum energy densities. Expanding Eq.~\ref{eq-initialrad} in the limit (generally satisfied when $M \ll 1$) where the tension is small compared to the energy splitting between the interior and exterior Hubble constants ($k_o^2 \ll H_F^2 - H_o^2$), we find this expression in terms of the variables for the thin-wall matching 
\begin{equation}\label{eq-smallbubapprox}
R^{i}_{o} \sim \frac{2 k_o}{\left[H_{F}^2 - H_{o}^2 \right]}.
\end{equation}

From the relation between the closed slicing coordinates and the embedding space coordinates, Eq.~\ref{eq-closedembedding}, we see that the bubble wall is parametrized by
\begin{equation}\label{obstrajectory}
\cos \eta = (1 - H_o^2 {R^{i}_{o}}^{2})^{1/2} \cos{T}.
\end{equation}
Similarly, the trajectory of the colliding bubble wall is given by
\begin{equation}\label{colltrajectory}
\cos (\eta - \eta_C) = - (1-H_{C}^2 {R^{i}_{C}}^{2})^{1/2} \cos{T}.
\end{equation}
with $R^{i}_{C}$ given by Eq.~\ref{eq-initialrad} with $o \rightarrow C$. The position and time of the collision $(T_{\rm coll}, \eta_{\rm coll})$ is obtained by finding the intersection between the curves Eq.~\ref{obstrajectory} and Eq.~\ref{colltrajectory}. 

We can also represent the bubble trajectories in terms of the ``hyperbolic foliation" of dS, specified by the embedding
\begin{eqnarray}
\label{Hds}
&& X_{0} = z \cosh \chi \\ \nonumber
&& X_{1} = H^{-1} \sqrt{1+H^2 z^2} \sin H x \\ \nonumber
&& X_{2} = z \sinh \chi \cos \phi \\ \nonumber
&& X_{3} = z \sinh \chi \sin \phi \\ \nonumber
&& X_{4} = H^{-1} \sqrt{1+H^2 z^2} \cos H x, 
\end{eqnarray}
which induces the metric Eq.~\ref{eq-dshyp}. Comparing Eq.~\ref{eq-closedembedding} with Eq.~\ref{Hds}, we can relate the global slicing to the hyperbolic slicing via
\begin{equation}\label{xtoeta}
{X_1\over X_4}=\tan Hx = \tan \eta \cos \theta,
\end{equation} 
where when $\theta=0$, then $\eta = Hx$. We can also see that the poles of the three-spheres of constant $T$ correspond to $Hx = 0, \pi$ (this is of course also true at any $\theta$). Again, comparing Eq.~\ref{eq-closedembedding} with Eq.~\ref{Hds}, we obtain for $z$:
\begin{equation}\label{ztoT}
z = H^{-1} \left[ \frac{\sin^2 \eta \cos^2 \theta  + \cos^2 \eta }{\cos^2 T}  -1  \right]^{1/2}.
\end{equation}
If $\theta=0$, then we have $z = H^{-1} \tan T$, which is also the value of $T$ at the poles for all $\theta$. The radius of the collision surface is therefore given in the hyperbolic coordinates by 
\begin{equation}\label{tcolltozc}
z_c = H_F^{-1} \tan T_{\rm coll},
\end{equation}
where $T_{\rm coll}$  is in turn a function of the separation of the nucleation centers, $\eta_C$.

Enumerating the parameters relevant for the kinematics, we have $(H_o, H_C, H_F, k_o, k_C)$ which are determined by the properties of the underlying potential landscape that is driving eternal inflation, and $\eta_C$ (or equivalently $z_c$), which will vary from collision to collision. However, all values of $\eta_C$ are not equally likely.  In Appendix~\ref{sec-appptol} we calculate the probability distribution of this variable, finding that for ``late time" collisions, it peaks near $\eta_C=\cos^{-1}(2/3)$ (although the exact position of the peak is dependent on the assumed cosmology inside of the bubble discussed in Appendix~\ref{bubblecosmo}), while for ``early time" collisions at a typical position, it peaks at $\eta_C = \pi/2$.  In neither case is the distribution narrow, but in both cases it falls to zero as $\eta_C \rightarrow \pi$ (for late-time collisions, not all values of $\eta_c$ are always possible, and so the distribution falls to zero much faster), which is the only regime where our results are sensitive to $\eta_C$.

\section{The post-collision spacetime}\label{postcollision}

\begin{figure*}[htb]
\includegraphics[width=13 cm]{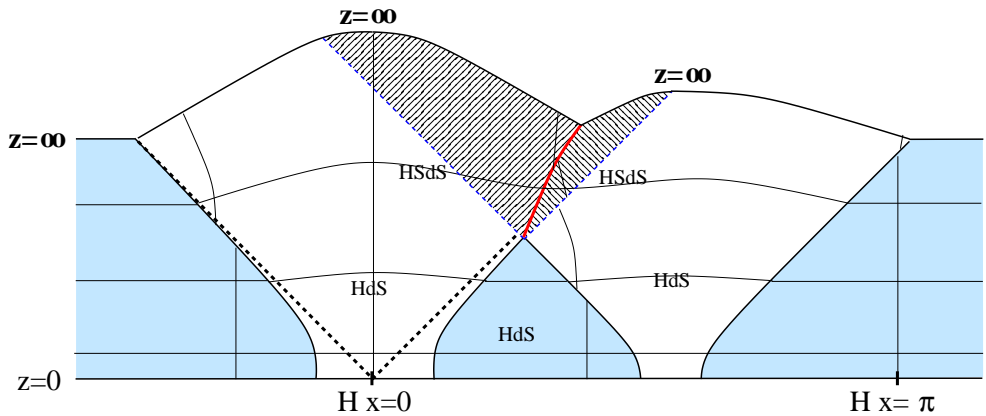}
\caption{The collision between two de Sitter bubbles displayed in the $(x,z)$ plane on a pseudo-conformal diagram (pseudo because not all of the points correspond to an $H_2$). Lines of constant $z$ and $x$ are drawn: note that $z$ is continuous across each of the junctions, but $x$ is discontinuous as required by the junction conditions. The post-collision domain wall is drawn as the solid red line, and the null shells as dashed blue lines. The causal future of the collision (shaded on the diagram) is Hyperbolic SdS, with a different mass parameter on each side of the post-collision domain wall.
  \label{fig-colldiag}
}
\end{figure*}

Assuming that the domain wall is thin, and that the vacuum energy on either side remains constant, then the model post-collision spacetime will consist of 5 separate regions of spacetime, each of which can be described by the general metrics discussed in Appendix~\ref{hyperspaces}, sewn smoothly (in the sense that the metric is continuous) together. A particular example of this is shown in Fig.~\ref{fig-colldiag}, which depicts the collision between two de Sitter bubbles. To the future of the collision, the metric is Hyperbolic Schwartzschild-de Sitter (HSdS), with a potentially different mass parameter on either side of the post-collision domain wall. The Israel junction conditions specify the procedure for matching across each of the null or timelike thin shells, and we must also be sure that all of the five regions are matched consistently across the collision surface (which is equivalent to requiring energy momentum conservation~\cite{Langlois:2001uq}).

In this section, we present the formalism necessary for performing these procedures in general. 
 
\subsection{Timelike domain walls}\label{timelikedom}
Here, we derive the junction conditions across a thin timelike domain wall of surface tension $\sigma_{oC}$ in a spacetime possessing hyperbolic symmetry. The metric on either side of the wall is of the form:
\begin{equation}\label{metric}
ds_{o,C}^{2} = - a_{o,C}(z)^{-1} dz^2 +  a_{o,C}(z) dx^2 + z^2 dH_{2}^{2}
\end{equation}
where for a general vacuum solution we have
\begin{equation}
\label{eq-aoc}
a_{o,C} = 1 - \frac{2 M_{o,C}}{z} + H^2_{o,C} z^2,
\end{equation}
and the subscripts $o,C$ specify the metric on the side of the observation bubble and colliding bubble respectively. The properties of such spacetimes are described in Appendix~\ref{hyperspaces}. 

The metric on the wall worldsheet is given by
\begin{equation}
ds_{3}^2 = -d\tau^2 + z(\tau)^2 d H_{2}^{2},
\end{equation}
where $\tau$ is the proper time of an observer attached to the wall.
Henceforth, $g_{\mu \nu}$ ($\mu, \nu = 0,1,2,3$) and $\gamma_{ab}$ ($a,b=0,1,2$) will denote the spacetime and worldsheet metrics respectively. The first junction condition requires that $z$ (the radius of the hyperboloid) matches across the wall; $x$ is in general discontinuous. We will specify the nucleation center inside of the observation bubble to be at $x=0$, which is equivalent to centering the observation bubble on the north pole of the background dS (see Eq.~\ref{xtoeta}).

The normal to the domain wall between the two colliding bubbles is found by requiring orthogonality ($g_{\mu \nu} n^{\nu} \partial_{a} x^{\mu} = 0$) and unit norm ($n^{\mu} n^{\nu} g_{\mu \nu} = 1$), yielding
\begin{eqnarray}\label{normal}
n_{z} &=& \dot{x} \\ \nonumber
n_{x} & =& - \dot{z}
\end{eqnarray}
where here and below the dot refers to a $\tau$-derivative. To eliminate the sign ambiguity in choosing $\tau$, we fix
\begin{equation}\label{dotx}
\dot{x} \equiv \frac{\beta_{o,C}}{a_{o,C}} = \frac{\sqrt{\dot{z}^2 - a_{o,C}}}{a_{o,C}},
\end{equation}
where $\beta_{o,C}$ is defined as a function of $z$ (as opposed to $z$ and $\dot{z}$ as it is here) in Eq.~\ref{beta} below. The non-zero components of the energy momentum tensor for the wall are assumed to be 
\begin{equation}
\label{eq-ktuv}
T_{a b} = -\sigma_{oC} \delta (z-z_{\rm wall}) \gamma_{a b}.
\end{equation}
Integrating Einstein's equations across the wall yields
\begin{equation}
K_{C \ b}^{a} - K_{o \ b}^{a} = -k_{oC} \delta_{b}^{a},
 \end{equation}
where $k_{oC} = 4 \pi \sigma_{oC}$ and $K_{ab}$ is the extrinsic curvature, given by
\begin{equation}\label{extrinsic}
K_{ab} = - \partial_{a} x^{\mu} \partial_{b} x^{\nu} D_{\nu} n_{\mu},
\end{equation}
where $D$ is the covariant derivative. The $\chi \chi$ component of the extrinsic curvature (which is all we require presently to find the equations of motion) yields
\begin{equation}
K_{\chi \chi} = \Gamma^{z}_{\chi \chi} n_{z} = z \beta 
\end{equation}
We finally obtain for the junction condition:
\begin{equation}
g^{\chi \chi} (K_{C \ \chi \chi}  - K_{o \ \chi \chi} ) = \frac{1}{z^2} (z \beta_{C} - z \beta_{o}) = -k_{oC},
\end{equation}
or:
\begin{equation}
\beta_{o} - \beta_{C} = k_{oC} z.
\end{equation}
Squaring this, solving for $\beta_{o,C}$, and using Eq.~\ref{dotx} gives:
\begin{equation}\label{beta}
\beta_{o,C} = \frac{a_{C} - a_{o} \pm k_{oC}^2 z^2}{2 k_{oC} z},
\end{equation}
where here and below the top sign refers to the ``o" subscript and the bottom to the ``C" subscript.

We can cast the junction conditions in an effective-potential form (see for example~\cite{Blau:1986cw}):
\begin{equation}\label{zeom}
\dot{z}^2 + V_{\rm eff} (z) = 1,
\end{equation}
where (using Eqs.~\ref{eq-aoc}, \ref{dotx}, \ref{beta}, and~\ref{zeom}) $V_{\rm eff}$ is given by:
\begin{eqnarray}\label{veff}
V_{\rm eff} &=& -\frac{1}{z^4} \frac{ \left( M_{o} - M_{C} \right)^2 }{k_{oC}^2} \\
&+& \frac{1}{z} \frac{ \left(H_{o}^2 - H_{C}^2 \right) \left( M_{o} - M_{C} \right) + k_{oC}^2 \left( M_{o} + M_{C} \right)}{k_{oC}^2} \nonumber \\
&-& z^2 \frac{ (H_C^2 + H_o^2 + k_{oC}^2 )^2 - 4 H_C^2 H_o^2 }{4 k_{oC}^2}. \nonumber
\end{eqnarray}
If we denote by $x_{o,C}$ the $x-$coordinate of the wall's trajectory (where the subscript denotes which side of the domain wall the trajectory is evaluated on), we can solve formally for $x_{o,C}(z)$, giving:
\begin{equation}\label{xoC(z)}
x_{o,C} = \int_{z_c}^{z} dz \frac{\beta_{o,C}}{a_{o,C} (1 - V_{\rm eff})^{1/2}}.
\end{equation}
This can be finite if there is a non-zero positive late time vacuum energy inside of the bubble (or if the solution is asymptotically timelike), even as $z \rightarrow \infty$ (owing to the finite range of $x$ in Hyperbolic de Sitter -- see Eq.~\ref{eq-dshyp}). Because of the causal structure of Hyperbolic de Sitter, it is impossible for the collision region to encompass all of future infinity inside of the bubble (the maximum range in $x$ reflects this fact). Given the parameters $(H_o, H_C, H_F, M_o, M_C, k_o, k_C, k_{oC})$, the only free parameter is the position $z_c$ of the collision, which is determined by the kinematics through Eq.~\ref{tcolltozc}.

\begin{figure}[htb]
\includegraphics[width=7cm]{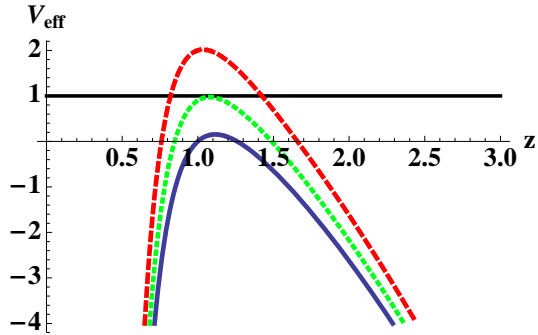}
\caption{The effective potential Eq.~\ref{veff} for a variety of parameters. The solid horizontal line at $V_{\rm eff} = 1$ denotes the ``energy" of the trajectory. Depending on the parameters in the potential, (i.e. if $V_{\rm eff} > 1$ for some range of $z$), there will be monotonic, bound and unbound trajectories in the presence of $V_{\rm eff}$. Only the monotonically increasing solutions are physical.
  \label{Veff}
}
\end{figure}

Returning to the effective potential Eq.~\ref{veff}, we see in Fig.~\ref{Veff} that there are three possible types of trajectories depending on the parameters of the potential: bound, unbound, and monotonic. Only solutions for which $z$ is monotonically increasing will be relevant for bubble collisions. Bound solutions must end with a singularity on both sides of the domain wall (at $z=0$, where Eq.~\ref{eq-aoc} blows up), which will in some cases be timelike (see Appendix~\ref{hyperspaces}). Unbound solutions that include a turning point must cross a killing horizon ($z$ must go spacelike in some region in order to go from decreasing to increasing), and therefore the collision spacetime will include a timelike singularity. These singularities are naked, and their formation would render the Cauchy problem to the future of the collision ill-defined. Fortunately, it is kinematically impossible to produce singularities in a collision~\cite{Hawking:1982ga,Wu:1984eda,Moss:1994pi} if the null energy condition is satisfied throughout the spacetime~\cite{Freivogel:2007fx,Chang:2007eq}.

The large-$z$ behavior is independent of the choice of mass parameters, and in this regime we obtain
\begin{equation}
\dot{z}^2 - c^2 z^2= 1
\end{equation}
where $c^2$ is the coefficient of the quadratic term in Eq.~\ref{veff}. When this constant is nonzero, the solution is exponential in $\tau$, $z \propto \exp\left( c \tau  \right)$, and elapsed proper time over the full range of $x_{o,C}$ is finite (as can be seen by integrating Eq.~\ref{dotx}) implying that the domain wall reaches future null infinity. If we allow for negative cosmological constants in the observation and colliding bubbles, then it is possible to have a tension for which $c=0$, given by
\begin{equation}\label{kcrit}
k_{\rm crit}^2 = - (H_{o} -H_{C})^2.
\end{equation}
For a real $k_{\rm crit}$, both $H_{o}^2$ and $H_{C}^2$ must be negative, or one must be zero and the other negative (recall that by our conventions, $H^2<0$ for AdS). The elapsed proper time along these trajectories is always infinite, implying that the domain wall reaches {\em timelike} future infinity, and they correspond to the BPS solutions of Ref~\cite{Freivogel:2007fx}.

\begin{figure}[htb]
\includegraphics[width=8.6cm]{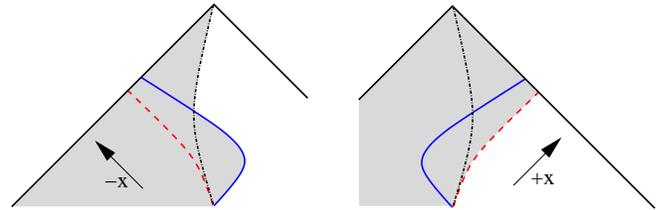}
\caption{A sketch of the possible asymptotic trajectories $x_o (z)$ inside of an observation bubble with zero cosmological constant. The shaded regions correspond to the possible regions that are on the observation bubble side of the domain wall. On the left, the asymptotic behavior of Eq.~\ref{assymtwo} is shown. In this case, the domain wall can be asymptotically null with (the solid blue line) or without (the dashed red line) a turning point, or timelike (the black dotted line). In the case where the wall is asymptotically null, all of future timelike infinity inside of the observation bubble is removed, and in the case where it is timelike, only a portion is removed. On the right, the asymptotic behavior Eq.~\ref{assymone} is shown. Again, the domain wall can be asymptotically null with or without a turning point, or timelike. In the case where the wall is asymptotically null, all of future timelike infinity inside of the observation bubble is preserved, and when it is timelike, only a portion is preserved. 
  \label{fig-trajectories}
}
\end{figure}

Another important feature of a trajectory is the sign of $\dot{x}$. This directly specifies the direction in which the domain wall travels, i.e. towards or away from the interior of the observation bubble. Our conventions dictate that when $\dot{x}<0$, the domain wall is moving into the observation bubble and when $\dot{x}>0$ it is moving into the colliding bubble. From Eq.~\ref{dotx}, since we require $z$ to be timelike everywhere along the trajectory, the sign of $\beta$ (Eq.~\ref{beta}) will fix the sign of $\dot{x}$. As $z \rightarrow \infty$, the sign of $\dot{x}$ on either side of the junction is therefore determined (via Eqs.~\ref{eq-aoc}, \ref{dotx}, and~\ref{beta}) by
\begin{eqnarray} \label{assymone}
\lim_{z \rightarrow \infty} \dot{x}_{o,C} &>& 0 \ : \ H_{C}^2 - H_{o}^2 \pm k_{oC}^2 > 0,  \\
\lim_{z \rightarrow \infty}  \dot{x}_{o,C} &<& 0 \ : \ H_{C}^2 - H_{o}^2 \pm k_{oC}^2 < 0, 
\label{assymtwo}
\end{eqnarray}
where the positive sign is taken when interested in finding asymptotics for $x_o$ and the negative sign is taken for $x_C$. In the case where $\dot{x}_{o,C} > 0$ the wall is moving into the colliding bubble and in the  case where $\dot{x}_{o,C} < 0$, it is moving into the observation bubble. This is the asymptotic behavior, but when there is a real positive root of $\beta_{o,C}=0$, there will be a sign change in $\beta_{o,C}=0$ located at
\begin{equation}\label{zbeta}
z_{\beta_{o,c}} = \left[\frac{2 (M_C - M_o) }{ H_C^2 - H_o^2 \pm k_{oC}^2 }\right]^{1/3},
\end{equation}
indicating that there is a turning point along the trajectory $x(z)$. In total, there are five qualitatively different trajectories to consider, as shown in Fig.~\ref{fig-trajectories}: those that go to $x \rightarrow x_{\rm max}$ (where in the example shown here $x_{\rm max} = \infty$ since the bubble interior is asymptotically flat) with or without a turning point, and trajectories which make it to $z \rightarrow \infty$ only after an infinite proper time (the critical solutions satisfying Eq.~\ref{kcrit}).

\subsection{Null domain walls}
To complete the problem, there are still two more junctions to consider across null surfaces, corresponding to the shells of radiation emitted from the collision event. Again, we require that the radii $z$ of the hyperbolae match across the junction. The null hypersurface can be specified by the vector $l^{\mu}$ tangent to the null geodesic generators of the shell and a null vector $n^{\mu}$ normalized such that $g_{\mu \nu} l^{\mu} n^{\nu} = -1$. We choose a basis where in the collision frame, the nonzero components of these vectors are given by
\begin{eqnarray}
l_z = 1&,& \ l_x = a^{-1} \nonumber \\
n_z = a/2 &,& \ n_x = -1 / 2
\end{eqnarray}
Assuming an energy momentum tensor of the form
\begin{equation}
T_{\mu \nu} = \sigma l_{\mu} l_{\nu} \delta ({\bf x}_{wall}),
\end{equation}
and integrating Einstein's equations across the wall, the junction condition is
\begin{equation}
k_1 - k_2 = 8 \pi \sigma,
\end{equation}
where we have chosen a convention such that the surface is moving from region 1 into region 2. Using Eq.~\ref{extrinsic} to calculate the components of the extrinsic curvature and tracing, this becomes
\begin{equation}
a_2 - a_1 = 8 \pi \sigma z.
\end{equation}
Since we assume that the cosmological constant remains the same across the null surface, then the junction condition will simply relate the mass parameters across the shell to the surface tension
\begin{equation}\label{energydensity}
M_1 - M_2 = 4 \pi \sigma z^2.
\end{equation}
From this relation, we see that the energy density falls like $z^{-2}$ as the shell propagates into the bubble. For the model collision spacetime described in Sec.~\ref{exactsolutions}, the mass parameter is non-zero only inside of the null shell, making $M_2=0$, and directly relating the energy density in the shell to $M_1$.

\subsection{Energy Conservation at the junction}

In this section we generalize the treatment of~\cite{Langlois:2001uq} to the hyperbolic metrics under consideration in the collision process (see also Ref.~\cite{Freivogel:2007fx}, whose conventions we follow). The boost angle between the rest frames of two colliding walls can be defined by
\begin{equation}
\label{eq-boostcosh}
g_{\mu \nu} U_{i}^{\mu}  U_{j}^{\nu} = \cosh{\pm \xi_{ij}}
\end{equation}
where the angle is negative when one wall is incoming (one of the initial state, pre-collision, walls) and the other is outgoing (one of the final state, post-collision, walls) and positive when both are incoming or outgoing, and $g_{\mu \nu}$ is the metric in the spacetime between the colliding walls. Performing a series of boosts between an arbitrary number of domain walls, we should find that upon coming back to the original frame, the total sum of the boost angles is zero:
\begin{equation}\label{conservation}
\sum_{i}^{n} \xi_{i\ i+1} = 0.
\end{equation}
This constraint can be written as energy or momentum conservation as seen in a particular frame~\cite{Langlois:2001uq}.

We will consider all domain walls to be timelike, and then take a limit for null shells. The four velocity of a timelike domain wall can be found by requiring $g_{\mu \nu} U^{\mu} U^{\nu} = -1$ and $g_{\mu \nu} n^{\mu} U^{\nu} = 0$ where $n^{\mu}$ is the normal constructed in Eq.~\ref{normal}, yielding:
\begin{equation}
\label{eq-wallnorm}
U_{x} = a \dot{x}, \ \ U_{z} = -\frac{\dot{z}}{a}. 
\end{equation}
The boost angle can be decomposed as follows. Using Eq.~\ref{eq-wallnorm} in Eq.~\ref{eq-boostcosh},
\begin{equation}\label{eq:xiij}
\xi_{ij} =  \pm \cosh^{-1} \left[ a \dot{x}_i \dot{x}_j - a^{-1}\dot{z}_i \dot{z}_j \right].
\end{equation}
where the boost angle is positive if walls $i$ and $j$ are both incoming or outgoing, and the boost angle is negative if one is incoming and the other outgoing. We now use Eq.~\ref{dotx}, where $\dot{x}_i$ and $\dot{x}_j$ could take opposite signs
\begin{eqnarray}\label{iioo}
\xi_{ij} &=& \pm \cosh^{-1} \left[ \pm \left(\frac{\dot{z}_{i}^{2} }{a } -1 \right)^{1/2}  \left( \frac{\dot{z}_{j}^{2} }{a } -1\right)^{1/2}  + \frac{\dot{z}_i}{a^{1/2}}  \frac{\dot{z}_j}{a^{1/2}}  \right] \nonumber \\
&=&   \pm \left[ \cosh^{-1} \left( \frac{\dot{z}_i}{a^{1/2}} \right) \pm \cosh^{-1} \left( \frac{\dot{z}_j}{a^{1/2}} \right) \right.
\end{eqnarray}
The overall sign is fixed as in Eq.~\ref{eq:xiij}, and the relative sign of the terms in parentheses is positive when wall $i$ and wall $j$ have the opposite sign of $\dot{x}$ and negative when both walls have the same sign of $\dot{x}$.

Note that the boost between wall rest frames has been decomposed into the sum of a boost from the rest frame of one wall to the ``rest frame" of the background spacetime (i.e., the frame defined by constant positions in the static coordinatization) and a boost from the rest frame of the background to the rest frame of the second wall. Because of this, the sum Eq.~\ref{conservation} can always be arranged as the sum of boosts between background rest frames,
\begin{equation}\label{sum2}
0 = \sum \left[\cosh^{-1} \left( \frac{\dot{z_i}}{\sqrt{a_{\alpha}}} \right)  - \cosh^{-1} \left( \frac{\dot{z_i}}{\sqrt{a_{\delta}}} \right) \right],
\end{equation}
where the wall labeled $i$ is entering region $\alpha$ from region $\delta$, and the sum is over all walls. We will sometimes find it useful to write the individual terms in this sum as
\begin{equation}
 \cosh^{-1} \left[ \frac{\dot{z_i}}{\sqrt{a_{\alpha}}} \right]  = \sinh^{-1} \left[ \frac{\beta_{\alpha}}{a_{\alpha}}\right]
\end{equation}
where the sign information is encapsulated in $\beta$, and one does not have to keep track of which wall is ingoing and which is outgoing.

The boost angle between two null shells, or a null and a timelike shell, is formally infinite. However, we can view it as the null-limit of the angle between timelike shells, where the condition Eq.~\ref{conservation} is maintained. This can be seen by noting how the boost angle in Eq.~\ref{iioo} has been decomposed: a divergent boost angle is required to go from one background rest frame to the null wall rest frame, but a divergent boost angle of opposite sign is required to go from the null wall rest frame to the background frame on the other side of the wall. Using the identity $\cosh^{-1} q = \log(q + \sqrt{q^2 - 1} )$, we can re-write each term in Eq.~\ref{sum2} as
\begin{eqnarray}
 \cosh^{-1} \left( \frac{ \dot{z_i} }{ \sqrt{a_{\alpha} } } \right)  &-& \cosh^{-1} \left( \frac{\dot{z_i}}{\sqrt{a_{\delta}}} \right) = \nonumber \\ \frac{1}{2} \log \left( \frac{a_{\delta}}{a_{\alpha}} \right)  &+& \log \left( \frac{ \dot{z}_i + \sqrt{\dot{z}_i - a_{\alpha} } }{ \dot{z}_i + \sqrt{\dot{z}_i - a_{\delta}} }\right).
\end{eqnarray}
The second term goes to zero in the limit of a null shell, and we are left only with the (finite) logarithm of the metric coefficients.\footnote{An interesting limit of the sum Eq.~\ref{sum2} occurs when all of the shells are null~\cite{Moss:1994pi}:
\begin{equation}
\prod_{\alpha} a_{\alpha} = \prod_{\delta} a_{\delta}
\end{equation}
where the product is over an arbitrary number of the metric coefficients in each term of the sum Eq.~\ref{sum2}.}

\section{Implications for bubble-collision observations}\label{detailedcoll}

As mentioned in the Introduction and discussed at length in AJS, for a bubble formed in a background spacetime of Hubble constant $H_F$, with other bubbles (of the same or different type) with nucleation rates (per unit 4-volume) $\lambda$, there are three regimes in which observers might expect to see collisions. 

If $\lambda \sim H_F^{4}$, then we are near the bound for bubble percolation, and {\em any} observer has a reasonable chance of seeing a collision, the angular scale (on an early constant-time surface) of which could be appreciable. In general, however, an exponentially suppressed nucleation rate $\lambda \ll H_F^{4}$ is expected~\footnote{In the case where gravitational effects are subdominant, and the thin-wall limit can be applied, the transition rate is set by $\lambda = A e^{-B}$, where A is a pre-factor and the bounce action scales like $B \propto \sigma^4 / (\Delta V)^3$~\cite{Coleman:1977py}, with $\Delta V$ the difference in the potential at the true and false minima. Significant fine-tuning of the potential is necessary to arrange $\sigma \sim (\Delta V)^{3/4}$ (because these scales are set by the properties of the barrier and the minima respectively) and thus $B\sim 1$. In the cases where gravity is important, the semiclassical approximation used to derive the Euclidean action breaks down before significant rates can be achieved.}. In this case there are still two regimes where collisions can be likely.  In ``late time" collisions, the observer's past lightcone encompasses many false-vacuum Hubble 4-volumes; but the effect is rather sensitive to the assumed cosmology inside of the bubble, as we discuss in Appendix~\ref{bubblecosmo}.\footnote{The estimate made by AJS, which did not account for an inflationary epoch inside of the bubble, is not appropriate for evaluating the expected number of collisions seen in our universe. We present a revised estimate in Appendix~\ref{bubblecosmo} which is closer to previous bounds~\cite{Gott:1984ps}.} Such bubbles have tiny observed angular size and are isotropically distributed over the sky. The second regime is that of ``early time" collisions which enter the past light cone of observers at $\xi_o \agt \lambda^{-1} H_{F}^{4}$ in the steady-state frame (where essentially all observers will be) at very early times. Each collision of this type covers nearly the full sky, and the distribution of such collisions is anisotropic. 

These conclusions relied on the approximation that the interior of the observation bubble remains undisturbed. However, these classes of observers and their associated picture of the collision events, could still be well-defined even after taking the effects of collisions on the interior of the observation bubble into account. For example, if $\lambda \sim H_F^{4}$ there may only be a modest boost  between the observation and collision frames. Therefore, we can imagine the observer at the origin in the collision frame. If the domain wall accelerates away from the observation bubble (as per the criterion given in Eq.~\ref{assymone}), these collisions stand a good chance of being (in the language of AJS~[AJS]) ``compatible" or ``perturbative" -- i.e. effecting the relevant portion of the observation bubble to a small or even perturbative degree, in a way that does not preclude observers.  The chances of a small effect are even better if there are circumstances in which, additionally, the domain wall does not have a turning point inside the observation bubble, (constraining $z_c > z_{\beta_{o}}$, with $z_{\beta_{o}} $ from Eq.~\ref{zbeta}). 

Early time collisions are large on the sky, but are potentially far more dangerous even for domain walls that accelerate away from the observation bubble.  This is because in the collision frame the observer is at $\xi_o \gg 1$, highly boosted with respect to the domain wall, radiation shell, and metric perturbations in the post-collision region.  An interesting question, then, is whether there are collision types that may be relatively benign even to these observers. In this section, we focus on assessing the detailed effects of the most mild types of collisions, calculating the size of the mass parameter $M_o$, and the degree to which the domain wall can be excluded from the observation bubble.

\subsection{Some like it mild}

The post-collision spacetime is fully specified by the set of parameters $(H_o, H_C, H_F, k_o, k_C,k_{oC},\eta_{C},M_{o},M_{C})$. We will assume that the Hubble constants and tensions are fixed by the potential landscape, but might vary significantly from collision to collision due to the variety of transitions that might exist. If $\eta_{C}$ is fixed at its most probable position (as derived in Appendix~\ref{sec-appptol}), then energy conservation forces a relation between $M_{o}$ and $M_{C}$. We will further assume that the initial radii of the observation and colliding bubbles are small compared to $H_F^{-1}$, which, from the discussion around Eq.~\ref{eq-initialrad} and Eq.~\ref{eq-smallbubapprox}, corresponds to tensions 
\begin{equation}
\label{eq-smallbub}
k_{o,C} \ll \sqrt{H_F^2 - H_{o,C}^2}.
\end{equation}
Although our construction is strictly valid only for bubbles inside of which the vacuum energy is held constant, the observation bubble must contain an epoch of slow-roll inflation and an epoch of late time acceleration (ie a changing vacuum energy) in order to describe our universe.\footnote{Note that for a viable model of open inflation, there must be additional mass scales in the scalar potential on top of the two discussed in Sec.~\ref{beforecollision}; see, e.g.~\cite{Linde:1995rv}} The inflationary $H_o$, which will be essentially constant over some range of open slicing time during slow roll, is most relevant for determining the evolution of the collision types we are considering. If the post-collision domain wall does not penetrate significantly into the observation bubble, the approximation of using the inflationary $H_o$ to determine its dynamics should be appropriate; even if it does, a lower effective $H_o$ will only cause the domain wall to accelerate away more quickly, so taking the inflationary value is conservative in this context.

Within these approximations, the kinematics will typically yield a value for $z$ at the collision which is of order $z_c \sim H_{F}^{-1}$ (this can be seen from the right panel of Fig.~\ref{fig-frames} and Eq.~\ref{tcolltozc}, and is rather robust as long as the separation between the nucleation centers does not approach $\eta_C \sim \pi$). Immediately, we obtain a constraint on the possible mass parameters $M_{o,C} \alt H_F $ since the collision must occur at a value of $z_c$ outside of the horizon (or a naked singularity is produced, a scenario that is not possible as discussed in Sec.~\ref{timelikedom}).  We can obtain a more precise estimate of the mass parameters by explicitly solving the condition for energy conservation, Eq.~\ref{conservation}, which is given by
\begin{widetext}
\begin{eqnarray}\label{eq-enconsreal}
0 &=&   \cosh^{-1} \left[ \frac{\sqrt{1 + z_c^2 / {R^{i}_{o}}^2  } }{\sqrt{ 1+ H_{F}^{2} z_c^2}} \right]  - \cosh^{-1} \left[ \frac{\sqrt{1 + z_c^2 / {R^{i}_{o}}^2  } }{\sqrt{ 1+ H_{o}^{2} z_c^2}} \right]  +  \cosh^{-1} \left[ \frac{\sqrt{1 + z_c^2 / {R^{i}_{C}}^2  } }{\sqrt{ 1+ H_{F}^{2} z_c^2}} \right] -  \cosh^{-1} \left[ \frac{\sqrt{1 + z_c^2 / {R^{i}_{C}}^2  } }{\sqrt{ 1+ H_{C}^{2} z_c^2}} \right]  \nonumber \\ 
&+& \sinh^{-1} \left[ \frac{\beta_o}{\sqrt{1 - \frac{2 M_o}{z_c} +H_o^2 z_c^2 } } \right] - \sinh^{-1} \left[ \frac{\beta_C}{\sqrt{1 - \frac{2 M_C}{z_c} +H_C^2 z_c^2 } } \right]  \\
&+& \frac{1}{2} \log \left[ \frac{(1+ H_{C}^{2} z_c^2)(1+ H_{o}^{2} z_c^2) }{ \left(  1 - \frac{2 M_C}{z_c} +H_C^2 z_c^2 \right) \left(  1 - \frac{2 M_o}{z_c} +H_o^2 z_c^2 \right)   } \right]. \nonumber
\end{eqnarray} 
\end{widetext}
It is possible to express this as an algabraic relation between $M_o$ and $M_C$ since the inverse hyperbolic functions are simply related to logarithms, however this will in general be an extremely complicated expression. Nevertheless, if the boost angles are all rather large (which is consistent with the small-bubble limit of Eq.~\ref{eq-smallbub}, as the bubble walls will be close to null at the collision) and $k_{oC}$ is parametrically smaller than $H_F$ (which helps, via Eq.~\ref{beta}, to guarantee the former condition), we can solve for $M_0$:
\begin{equation}\label{Mo}
M_o \simeq \frac{ (H_F^2 - H_C^2) (1+H_{o}^2 z_c^2) }{2 (1 + H_F^2 z_c^2 )} z_c^3. 
\end{equation}
There are corrections involving $M_C$, but restricting our attention to solutions that do not have a turning point in $x(z)$, $M_C$ lies in the range \begin{equation}
0 \leq M_C \leq M_o + \frac{z_c^3}{2} \left(H_C^2 - H_o^2 - k_{oC}^2 \right),
\end{equation}
over which the corrections are negligible. We have checked the behavior of Eq.~\ref{Mo} against exact solutions  to Eq.~\ref{eq-enconsreal} for a wide range of parameters consistent with our approximations, and found excellent agreement. In this regime, we therefore conclude that $M_o$ is never zero and will typically have a scale set by $z_c$. Significant corrections to Eq.~\ref{Mo} are introduced when $k_{oC}$ or $z_c^{-1}$ become large compared to $H_F$, since we can no longer neglect many of the terms in Eq.~\ref{eq-enconsreal}. The energy density in the null shell entering the observation bubble at the time of the collision, using Eq.~\ref{energydensity} with $z_c \sim H_F^{-1}$, will typically be of order (replacing factors of $m_p$) $\sigma (z_c) \sim 0.1 H_F m_p^2$.

An additional feature that would be desirable for determining the extent to which a collision is perturbative is how quickly the post-collision domain wall accelerates away from the interior of the observation bubble. Asymptotically, the trajectory $x(z)$ will be independent of the mass parameters, and integrating Eq.~\ref{xoC(z)} at large $z$ we obtain:
\begin{eqnarray}
x_{o} (z) &=& \left[ 1+ \left( \frac{2 H_o k_{oC}}{H_C^2 - H_o^2+k_{oC}^2 } \right) \right]^{-1/2} \nonumber \\ &\times& H_{o}^{-1} \tan^{-1} (H_{o} [z-z_0]) + x(z_0).
\end{eqnarray}
Because a null line in HdS is parametrized by $x = H_o^{-1} \tan^{-1} [H_{o} (z - z_0 )]  + x(z_0)$, we see that the wall is closest to being null when
\begin{equation}
1 + \frac{2 k_{oC}}{H_{o}} - \left( \frac{k_{oC}}{H_o}\right)^2  \ll \left( \frac{H_{c}}{H_{o}} \right)^2,
\end{equation}
which can be arranged by having a hierarchy $k_{oC} \ll H_o \ll H_C$. Studying the post-collision geometry more carefully using numerical solutions (taking into account the effect of the mass parameters in the vicinity of the collision), it is even possible to adjust the parameters such that the domain wall {\em never} enters the future light cone of the observation bubble's nucleation center.  (This depends on $z_c \sim H_F^{-1}$, but  this is true unless $\eta_c\simeq0$ or $\eta_c\simeq \pi$, which is in turn unlikely, as discussed in Sec.~\ref{beforecollision}.)

\subsection{Observer survival}

Having assessed the effects of a collision on the interior of the observation bubble, we return to the relationship between the collision and observation frames. All collision spacetimes having asymptotic behavior given by Eq.~\ref{assymone} and no turning point contain timelike geodesics that always remain inside of the observation bubble, and so the observation frame is well-defined for observers following these. In a bubble interior unaffected by collisions, observers located at various fixed $\xi_o$ positions for all $\tau_o$ can be defined via a congruence of timelike trajectories boosted with respect to the collision frame and passing through the nucleation center.  The observers with larger boost parameters are ``further" from the origin defining the observation frame, and this relationship is precise in the case where the interior of the observation bubble can be covered by an open FLRW patch.

\begin{figure}[htb]
\includegraphics[width=8.cm]{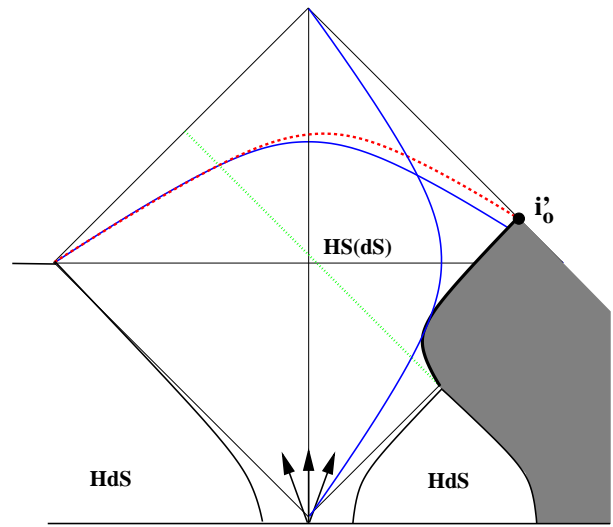}
\caption{ The FLRW patch inside of an observation bubble with zero asymptotic vacuum energy, after a collision. The congruence of boosted trajectories passing through the nucleation center can be used to define the radial position $\xi_o$ of various observers. In this example, the solid blue line shows one such trajectory which intersects the post-collision domain wall; trajectories with a larger boost will also intersect the domain wall. These are the trajectories that would have been followed by observers who see early time collisions (early because of the time dilation due to the large boost). The spacelike solid blue line denotes a surface of constant density in the unperturbed collision bubble. The red dotted line denotes a hypothetical distortion of this constant density surfaces due to the presence of the collision, in which the spacelike surface is still infinite in extent, with a spacelike infinity at $i_o'$ in the diagram.
  \label{fig-obsframe}
}
\end{figure}

Including the effects of collisions, such boosted trajectories are still well-defined, but when the post-collision domain wall enters the observation bubble, as shown in Fig.~\ref{fig-obsframe}, not all of them will reach future timelike infinity. The early-time collisions of AJS exist only for observers in this congruence which are extremely boosted with respect to the Steady State frame, and generally greatly boosted with respect to the collision frame as well. We might therefore worry that there is no such set of observers in the presence of collisions. 

What do our solutions say about this?  In the case where the domain wall enters the future lightcone of the observation bubble nucleation center but accelerates asymptotically away to intersect future null infinity (as depicted in Fig.~\ref{fig-obsframe}), it seems reasonable to expect infinite spacelike slices of constant field to exist in the collision region.  These will not be the same as the original constant-field surfaces (and will be joined to those surfaces outside of the collision region by a region in which the constant-field surfaces go timelike so that spacelike slices must have field inhomogeneities there); but there seems no reason to neglect the infinitely many observers in this region, all of whom see the bubble collision.

Even better, however, is the class of solutions found where the post-collision domain wall does {\em not enter} the light cone. Here, all of the observers in the congruence appear to make it to timelike infinity just as in the case where the bubble interior remains undisturbed. In this case, we might define what is meant by a translation inside the collision region as taking a boosted trajectory to an unboosted one, in exact analogy with the case where the observation bubble can be foliated by an FLRW patch. This suggests that even including the effect of the large boost necessary to go into the observer frame for them, the set of observers that record early-time collisions in this case may remain intact. This is because we can extend the null cone bounding the open slicing region of the observation bubble to define a null cone that is never crossed by the domain wall, and the lack of such intersections is boost-invariant.

\subsection{Potential observables}

Having established that early-time and late-time collisions, as well as the nearly percolating collisions, appear to allow observers to their future in at least some cases, we now discuss the potential observable signatures from such events, though it is not the focus of the present paper. Generically, the collision will form a disc on the observer's sky, the boundary of which corresponds to the intersection of the null shell with some spacelike hypersurface to the past that represents the cosmological time from which the observer is receiving the radiation seen. Therefore, one important signature of a single bubble collision (or multiple bubble collisions from the same direction) is azimuthal symmetry about some direction on the sky (as previously noted in~\cite{Aguirre:2007gy}). The angular size of this disc will typically be small for late time collisions, of order $\pi/4$ for the nearly percolating collisions (as discussed in Appendix~\ref{bubblecosmo}), and nearly $2 \pi$ for early time collisions.

What sort of effects might a collision produce inside the disk? As discussed in the text, although $z$ is continuous across the null shell, $x$ is not. There is thus an effective ``time dilation" in the future light cone of the collision, and as we follow the null surface towards increasing $z$, integrating $ds^2=0$ in Eq.~\ref{eq-aoc} shows that the discontinuity in $x$ is log divergent. If we imagine a field (which we are assuming here does not significantly backreact on the geometry, as say during slow-roll where the vacuum energy is nearly constant) inside the bubble rolling down from its (potentially different) value at the location of the post-collision and observation bubble domain wall, then the field will be effectively desynchronized across the null interface. The patch of the sky containing the collision will therefore have retarded field evolution as compared to the field history at other angles, or possibly correspond to an entirely different history of field evolution altogether. Because of this discontinuity, field gradients are also likely be very large along the boundary of the affected region on the sky. Several other possible observational signatures have been discussed in some detail in Ref.~\cite{Chang:2007eq}. Any of these types of signal are likely to be observable only when the angular scale encompassed by the collision is rather large, as in the nearly percolating and early time collision types. 

\section{Conclusions}\label{conclusions}

In a previous work (AJS~\cite{Aguirre:2007an}) we investigated the circumstances under which an observer in a ``bubble universe", formed via a first-order phase transition in the inflaton field, could have the impacts of other such bubbles within that observer's past lightcone. Here, we have computed exact solutions modeling bubble collisions in the limit of a perfectly thin wall and strictly constant vacuum energy.  While this simplified model leaves a number of questions unanswered, our calculations provide a number of useful results, several of them quite favorable for the prospect of observing bubble collisions. Our chief conclusions are:

\begin{itemize}

\item Using the symmetries of the bubbles and their collisions, as well as the ability to boost between frames in a way that is largely defined in terms of boosts in an embedding Minkowski space, the model bubble collision can be completely specified using ten parameters.  Six of these (vacuum energies of the bubbles and false vacuum, and tensions in the bubble walls and in the domain wall connecting them) are determined by the inflaton potential. This leaves a boost parameter, the invariant separation $\eta_C$ between bubble nucleation centers, and two mass parameters.  In Appendix~\ref{sec-appptol}, we calculate the distribution of $\eta_C$, finding that for a given class of observers it peaks at a preferred separation, which may then be assumed as generic for bubble collisions of that class. The distribution of boost parameters was effectively derived in AJS.
 
\item Within this construction, we have found that there are many solutions in which the observation frame is well defined, i.e. where the domain wall between the colliding bubbles accelerates away from, and does not impact, the observer. These solutions are specified by the condition Eq.~\ref{assymone}. 

\item Collisions may come in to the observer's past lightcone at any cosmological time if the nucleation rate is rather high (of order the rate necessary for bubble percolation).  If it is somewhat lower a ``late"-time observer may or may not see bubbles, depending on the assumed cosmology inside of the observation bubble, as discussed in Appendix~\ref{bubblecosmo}. In either case, we have provided criteria for whether the post-collision domain wall (if one exists) accelerates toward or away from such an observer.  If towards, the wall impact is very likely fatal, so combined with the results of AJS, this could be used to constrain (via our observed continued survival) models with a relatively high nucleation rate (given by Eq.~\ref{eq-lambdabound}). If away, potentially observable signatures of the bubbles could be used to rule out {\em or} confirm models with similarly high nucleation rates.

\item  Bubbles that enter the observer's lightcone at early times might be seen by essentially all observers might see -- even for a tiny nucleation rate -- provided those observers survive the collisions. The key question is to what degree the domain wall penetrates the observation bubble, and how large the gravitational distortions are. We have provided two results regarding this issue. First, we have shown that it is possible to construct solutions where the post-collision domain wall separating the interiors of two colliding bubbles monotonically moves away from the interior of the observation bubble, and even solutions for which the domain wall never enters the null cone emanating from the nucleation center of the observation bubble. Second, we have derived a simple scaling relation, Eq.~\ref{Mo}, for the mass parameter describing the distortion of the geometry to the future of the collision event. This distortion cannot be made arbitrarily small in this class of collision spacetimes, and will generally correspond to a shell of radiation which has an energy density set by the false vacuum Hubble parameter.
\end{itemize}

There are several clear directions for further research on the problem.  First, numerical simulations of bubble collisions can be used to both test the validity of the approximations used here, and also assess the detailed dynamics of the collision and the post-collision region.  Second, our results suggest that some collisions might be mild for {\em all} bubble observers, even those highly boosted with respect to the steady-state frame, for whom early-time, large-angular-scale collisions should be common.  A perturbative treatment of an observation bubble with a fluctuation on the wall can be used to both assess whether this is indeed the case, and also potentially translate the collision effects into CMB and other observables.  Both approaches are underway and should provide vital insight into whether we might have direct observational evidence that our universe is undergoing eternal inflation. 

\begin{acknowledgments}
The authors wish to thank T. Banks, R. Bousso, B. Freivogel, and S. Shenker  for helpful discussions as well as Marty Tysanner for useful comments on the draft. MJ acknowledges support from the Moore Center for Theoretical Cosmology at Caltech. AA was partially supported by a ``Foundational Questions in Physics and Cosmology" grant from the Templeton  Foundation during the course of this work. 
\end{acknowledgments}

\begin{appendix}
\section{Hyperbolic spacetimes}\label{hyperspaces}
In this appendix, we discuss some of the properties of metrics possessing hyperbolic symmetry. The general form of such metrics in the presence of a cosmological constant (using a version of Birkhoff's theorem~\cite{Wu:1984eda}) is
\begin{equation}
ds^{2} = - a(z)^{-1} dz^2 +  a(z) dx^2 + z^2 dH_{2}^{2}
\end{equation}
where $dH_{2}^{2}$ is given by
\begin{equation}
dH_{2}^{2} = d \chi^2 + \sinh^2 \chi d\phi^2 ,
\end{equation}
and in general we have
\begin{equation}
a = 1 - \frac{2 M}{z} + H^2 z^2 .
\end{equation}
One important consequence of this close relationship with spherically symmetric spacetimes is that no gravitational radiation can be produced if the O(2,1) invariance is maintained~\cite{Wu:1984eda}.

\begin{figure*}
\includegraphics[width=12cm]{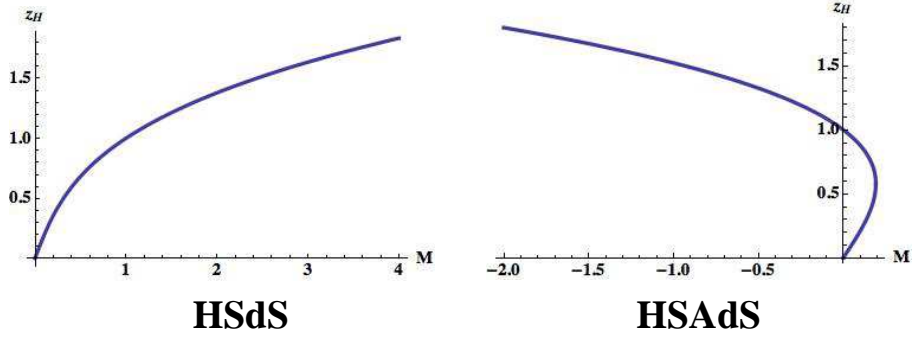}
\caption{The location of the horizon(s) in Hyperbolic SdS (left) and Hyperbolic SAdS (right) as a function of mass (with both quantities scaled to $H^{-1}$).
  \label{fig-horizons}
}
\end{figure*}

Solving $a=0$ for the location of any killing horizons that might exist, there are two qualitatively different cases. When $H^2 \geq 0$, there is one horizon as shown in the left panel of Fig.~\ref{fig-horizons}, which grows with $M$. When $H^2 < 0$, the situation is different: there are $z=$const. killing horizons at positive $z$ for both positive and negative mass parameter as shown in the right panel of Fig.~\ref{fig-horizons}. There is both an inner and outer horizon for $M > 0$, except in the ``extremal" case where $M= (3 \sqrt{3} |H|)^{-1}$ and the two horizons are degenerate (for more details, see Ref.~\cite{Freivogel:2007fx}). When $M<0$, there is only one horizon.

\begin{figure}
\includegraphics[width=9cm]{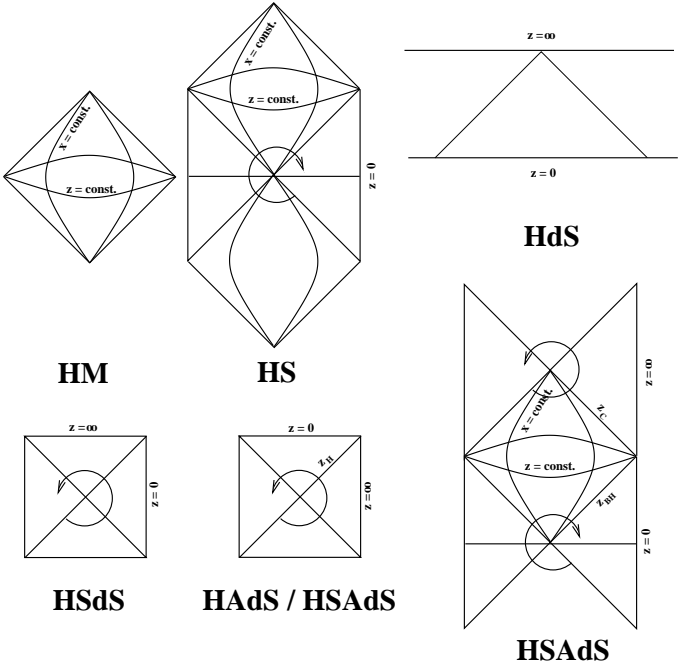}
\caption{The conformal structure of all spacetimes with a mass parameter and cosmological constant possessing hyperbolic symmetry including: Hyperbolic Minkowski (HM), Hyperbolic Schwarzschild (HS), Hyperbolic de Sitter (HdS), Hyperbolic Schwarzschild de Sitter (HSdS), Hyperbolic Anti de Sitter (HAdS), and Hyperbolic Schwarzschild Anti de Sitter (HSAdS). Each point on the diagram represents an $H_2$. All singularities are timelike, except for the case of SAdS with $M<0$. The circulating arrows denote the direction of increasing $x$ on each diagram.
\label{fig-hyperbolicspaces}
}
\end{figure}

Returning to the metric, we see that for $M\neq 0$ there is a curvature singularity at $z=0$. This singularity is timelike in all cases except when $H^2 < 0$ and $M < 0$, where it is spacelike. We can represent the global structure of these spacetimes by drawing their hyperbolic conformal diagrams (each point on the diagram corresponds to an $H_2$) as shown in Fig.~\ref{fig-hyperbolicspaces}. These conformal diagrams can be obtained from the well-known causal structure of spherically symmetric spacetimes with a cosmological constant (see eg~\cite{Lake:1977ui}) by taking $H^2 \rightarrow - H^2$ and rotating the diagram by 90 degrees~\cite{Wu:1984eda} (taking spacelike surfaces into timelike surfaces).

A useful tool for constructing the global structure of collision spacetimes is the sign of $\dot{x}$ (see Eq.~\ref{dotx}) along a domain wall. In spacetimes with a horizon, the direction of increasing $x$ changes (with surfaces of constant $x$ going from spacelike to timelike or vice verse) as indicated by the arrows in Fig.~\ref{fig-hyperbolicspaces}. A trajectory drawn on the conformal diagram must pass through the appropriate regions, for example, if we found a solution in $HS$ with $\dot{x}>0$ and $z<z_H$ for some range in $z$, then the trajectory must pass through the left wedge. The direction of the outward normal is fixed by the choice of sign in Eq.~\ref{dotx}, and in our conventions points towards larger $|x|$. 

\section{Distribution of nucleation centers}
\label{sec-appptol}

We saw in Sec.~\ref{beforecollision} that the de Sitter invariant distance between nucleation centers is the sole kinematical variable in a collision between two bubbles that is not fixed by the micro-physics. In~\cite{Carvalho:2002fc}, a probability distribution for the dS-invariant distance between collision centers in the steady-state frame was derived. We examine this distribution for bubbles that satisfy the additional constraint of nucleating inside the past light cone of an observer, which is clearly the appropriate measure for potentially observable collisions. We will find that the distribution is peaked in different positions for the early- and late-time collision types. This most probable separation of the nucleation centers can then be used to specify the kinematics for a typical collision event.

We begin with the differential number of bubbles nucleated in a parcel of 4-volume from AJS (which contains a similar calculation and to which we refer the reader for more details on the method):
\begin{equation} \label{eq-volelement}
dN = \lambda dV_4 = \lambda H_{F}^{-4} \frac{\sin^2 \eta_n}{\cos^4 T_n} dT_n d \eta_n d(\cos \theta_n) d\phi_n.
\end{equation}
If the nucleation rate is sufficiently small (if we are in the regime of eternal inflation~[AJS]), then we can neglect corrections due to the volume removed by individual bubble nucleation events~\cite{Carvalho:2002fc}. We will also make the approximation in this section that the observation bubble nucleates with zero radius (so that the bubble wall is a light cone; this is accurate under the conditions given by Eq.~\ref{eq-smallbubapprox}), and that the observation bubble interior is unaffected by collisions.

Using the analysis of Sec.~\ref{beforecollision}, we note that a measure of the proper distance between the nucleation center of the observation bubble and a spacelike separated point is just its position $(\eta_C, T_C = 0)$ in the collision frame. Returning to the observation frame leaves the quantity
\begin{equation}
I = \frac{\cos \eta}{\cos T} = \cos \eta_C
\end{equation}
invariant since $X_4$ (see Eq.~\ref{eq-closedembedding}) does not change upon transforming between the observation and collision frames in this case. Changing variables in Eq.~\ref{eq-volelement} from $\eta$ to $ \eta_C $ then yields
\begin{eqnarray}
dN = - \lambda H_{F}^{-4} &\sin& \eta_C \frac{\left( 1 - \cos^2 \eta_C \cos^2 T \right)^{1/2}}{\cos^3 T} \nonumber \\ &\times& dT d\eta_C d(\cos \theta) d\phi.
\end{eqnarray}
Integrating $T$ along a surface of constant $\eta_C$ gives the differential number of bubbles centered on an angular position $(\theta, \phi)$ seen by the observer which nucleated at a de Sitter invariant distance $\eta_C$ from the observation bubble:
\begin{eqnarray} \label{eq-sepdist}
\frac{dN}{d \eta_C d(\cos \theta) d\phi} &=& - \lambda H_{F}^{-4} \sin \eta_C \\ &\times& \int_{T_i}^{T_F} dT \frac{\left( 1 - \cos^2 \eta_C \cos^2 T \right)^{1/2}}{\cos^3 T}. \nonumber
\end{eqnarray}
The lower limit of integration is determined by the intersection of the initial value surface with the surfaces of constant $\eta_C$. Since we are working in the observation frame, the initial value surface will in general have angular dependence parametrized by~[AJS]:
\begin{equation}\label{boostedinitialvalue}
\sin T = - \left(\frac{\cos \eta}{\gamma_o} + \beta_o \sin \eta \cos \theta \right).
\end{equation}
The upper limit of integration is given by the intersection between the past light cone of the observer and the surfaces of constant $\eta_C$. The past light cone of the observer is specified by $T_{co}$, the global slicing time at which the light cone intersects the observation bubble wall. Matching the interior FLRW across the wall yields~[AJS]  
\begin{equation}\label{Tco}
T_{\rm co}=\arctan \left[ H_F \lim_{\tau \rightarrow 0} a(\tau) \sinh \left( \int_{\tau}^{\tau_{\rm o}}d\tau/a(\tau) \right) \right].
\end{equation}

\begin{figure}[htb]
\includegraphics[width=8cm]{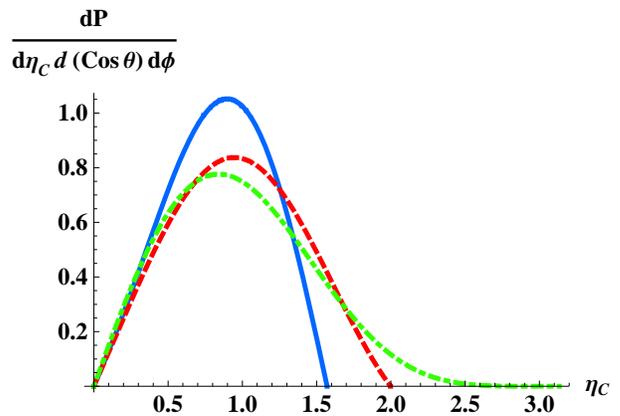}
\caption{The normalized distribution of bubble separations in the collision frame for late-time small-scale collisions for $T_{\rm co} = (\pi/4, 3 \pi / 8, \pi / 2)$ (blue solid, red dashed, green dot-dashed).
  \label{fig-latetimesep}
}
\end{figure}

Integrating Eq.~\ref{eq-sepdist}, we are interested in a variety of limits. Sampling observers at open-slicing position $\xi_o = 0$ will yield the separation distribution for late-time collisions only. As we discuss in Appendix~\ref{bubblecosmo}, the value of $T_{\rm co}$ corresponding to a given observer depends on the assumed cosmology inside of the bubble. If there is a non-zero late-time vacuum energy inside of the bubble, then there will be a maximum value of $T_{co} < \pi / 2$. Sampling observers at large-$\xi_o$ and small $\tau_o$ ($T_{\rm co} \rightarrow \pi / 2$) will yield the separation distribution for early-time collisions. Because nearly all collisions in this frame come at early cosmological time, this distribution is relatively insensitive to the assumed cosmology inside of the bubble. Starting with the late-time collisions, since the initial value surface has no angular dependence, the distribution is isotropic. The normalized distribution (the probability of a given separation given that nucleation events do occur) as a function of $\eta_C$ for various $T_{\rm co}$ is shown in Fig.~\ref{fig-latetimesep}. As $T_{\rm co} \rightarrow \pi / 2$, corresponding to an observer at $\tau_o \rightarrow \infty$ in a universe with no late-time vacuum energy, the distribution approaches the result of Ref.~\cite{Carvalho:2002fc}: 
\begin{equation}
\lim_{T_{\rm co} \rightarrow \pi / 2} \frac{dP_{\rm late} }{d \eta_C d(\cos \theta) d\phi} = \frac{3}{8} \sin \eta_C \left( 1 + \cos \eta_C \right)^2
\end{equation}
The maximum of this distribution is located at $\eta_C = \cos^{-1} (2/3)$. Note that unless $T_{\rm co} = \pi / 2$, not all invariant separations can be sampled by the observer. When there is a late-time vacuum energy inside of the bubble, since $T_{\rm co}$ asymptotes to a value less than $\pi / 2$, observers at rest with respect to the steady state frame will only see collisions at a separation less than some value determined by the precise cosmology inside of the bubble, as described below in Appendix~\ref{bubblecosmo}.

\begin{figure}[htb]
\includegraphics[width=8cm]{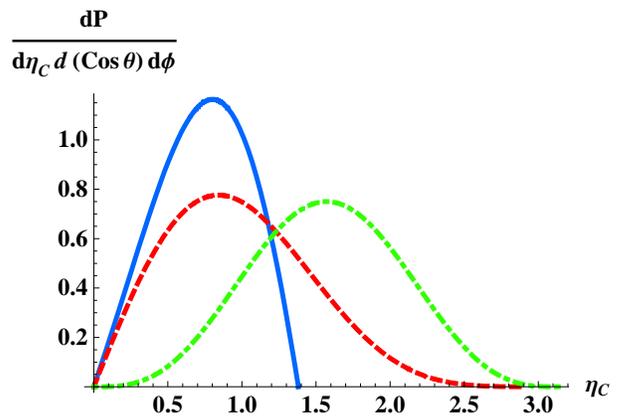}
\caption{The distribution of bubble separations in the collision frame for early-time large-scale collisions at $T_{\rm co} = \pi / 8$ and $\theta_o = 0$ for $\xi_o = (.5, 5 ,50)$ (blue solid, red dashed, green dot-dashed).
  \label{fig-earlytimesep}
}
\end{figure}

\begin{figure}[htb]
\includegraphics[width=8cm]{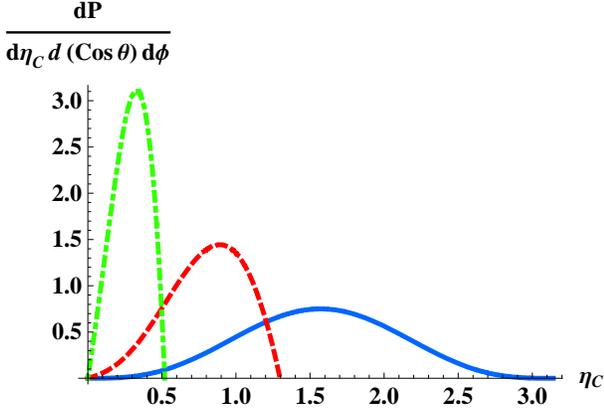}
\caption{The distribution of bubble separations in the collision frame for early-time large-scale collisions at $T_{\rm co} = \pi / 8$ and $\xi_o = 50$ for $\theta_o = (0,1, 2)$ (blue solid, red dashed, green dot-dashed).
  \label{fig-earlyangle}
}
\end{figure}

The normalized distribution for the early-time large-scale collisions is shown for various $\xi_o$ in Fig.~\ref{fig-earlytimesep}. At large-$\xi_o$, the maximum approaches $\eta_C = \pi / 2$ (at $\theta_o=0$), which is different than the most probable separation for the late-time small-scale collisions. In addition, the distribution will have angular dependence due to the distorted initial value surface. This is shown in Fig.~\ref{fig-earlyangle}.

\section{Cosmology and hat sizing of bubbles}\label{bubblecosmo}
The number of collisions in an observer's past light cone is determined by the quantity $T_{\rm co}$, Eq.~\ref{Tco} , which is the closed slicing time (as defined by the background eternally inflating de Sitter space) at the intersection of the observer's past light cone and the bubble wall. In AJS, this quantity was calculated under the assumption that the bubble interior was pure vacuum. However, including a more realistic cosmology inside of the bubble is clearly in order both to more precisely determine the expected number of collisions in our past light cone, and to determine the conformal structure of an arbitrary bubble universe (i.e., determine the size of the ``hat" on a conformal diagram, as we describe below). In this appendix we find the expected number of late-time collisions for an arbitrary cosmology inside of the bubble. (We also note the paper of Ref.~\cite{Gott:1984ps}, previously unknown to us, which calculated the expected number of late time collisions; our calculation is a generalization of this.) We find (consistent with the results of~\cite{Gott:1984ps}) that inflation inside the bubble is crucial, and that the assumption made by AJS that the bubble cosmological time determines the expected number of late-time collisions is incorrect. Instead, the expected number of collisions at asymptotically late times for an arbitrary cosmology is generally determined both by the vacuum energy on either side of the bubble wall and the minimal value of the density parameter $\Omega$; for our observed universe the latter dependence is negligible. 
Given this, we provide a bound on the nucleation rate $\lambda$ necessary to see (or not see) late-time bubble collisions, and comment on the angular scale that such collisions would take up on the sky. 

The cosmological evolution inside of the future light cone of the nucleation center can be described by the metric Eq.~\ref{eq-openfrw} with spacelike constant density hyperboloids
\begin{equation}
ds^2 = -d\tau^2 + a(\tau)^2 \left[ d \xi^2 + \sinh^2 \xi d\Omega_2^2 \right].
\end{equation}
We will assume the following evolution of the scale factor
\begin{eqnarray}
\label{eq-epochs}
a &=& a_I  = H_{I}^{-1} \sinh (H_I \tau), \ \  0 < \tau < \tau_I \nonumber \\
a &=& a_{w}, \ \ \tau_I < \tau < \tau_{\Lambda} \nonumber \\
a & = & a_{\Lambda} = a(\tau_{\Lambda}) \exp[ H_{\Lambda} (\tau-\tau_{\Lambda})], \ \ \tau_{\Lambda} < \tau < \infty.  
\end{eqnarray}
That is, we have a period of curvature domination (for $\tau \alt H_{I}^{-1}$), followed by a period of inflation at a scale fixed by $H_I$. After inflation end at time $\tau_I$, we allow for any number of epochs dominated by various components of the energy density characterized by an equation of state $p = w \rho$ (with an instantaneous transition between each epoch), followed by late-time vacuum energy domination. The current epoch in our universe will be denoted by $\tau_o$. 

Decomposing the integral in Eq.~\ref{Tco} into terms for the pre- and post-inflationary evolution
\begin{equation}
 \int_{\tau}^{\tau_{\rm o}}d\tau/a(\tau) =  \int_{\tau}^{\tau_{\rm I}}d\tau/a_{I}(\tau) +  \int_{\tau_I}^{\tau_{\rm o}}d\tau/a (\tau),
\end{equation}
and expanding $T_{co}$, we obtain
\begin{eqnarray}
\label{eq-tco}
T_{\rm co} &=& \arctan \left[ H_F  \lim_{\tau \rightarrow 0} a (\tau) \right. \\
 && \left[ \sinh \left( \int_{\tau}^{\tau_{I}}d\tau/a_I (\tau) \right) \cosh \left( \int_{\tau_I}^{\tau_{\rm o}}d\tau/a (\tau) \right)  \right. \nonumber \\
 && \left. \left. +  \cosh \left( \int_{\tau}^{\tau_{I}}d\tau/a_I (\tau) \right) \sinh \left( \int_{\tau_I}^{\tau_{\rm o}}d\tau/a (\tau) \right) \right]  \right] \nonumber .
\end{eqnarray}
Taking the limit as $\tau \rightarrow 0$ will only involve the terms depending on $a_I$, and we have that
\begin{eqnarray}
\lim_{\tau \rightarrow 0} && a_I  \sinh \left( \int_{\tau}^{\tau_{\rm I}} d\tau/a_I (\tau) \right) = H_{I}^{-1} \tanh(H_I \tau_I / 2) \nonumber \\
\lim_{\tau \rightarrow 0} && a_I  \cosh \left( \int_{\tau}^{\tau_{\rm I}}d\tau/a_I (\tau) \right) = H_{I}^{-1} \tanh(H_I \tau_I / 2). \nonumber
\end{eqnarray}
Using this in Eq.~\ref{eq-tco} gives
\begin{equation} \label{Tcosimp}
T_{\rm co} = \arctan \left[ \frac{H_F}{H_I} \tanh \left( \frac{H_I \tau_I}{2} \right) \exp \left( \int_{\tau_I}^{\tau_{\rm o}}d\tau/a (\tau) \right)\right].
\end{equation}

The integral in the argument of the exponential (which we will denote by ${\cal I}$) can be split into two terms, representing the two qualitatively different post-inflationary epochs of Eq.~\ref{eq-epochs}
\be
\label{eq-ii}
{\cal I} = \int_{\tau_{I}}^{\tau_{\Lambda}} \frac{d\tau}{a_w} + \int_{\tau_{\Lambda}}^{\tau_o} \frac{d\tau}{a_{\Lambda}}.
\ee
Concentrating on the first integral, let $\Xi (\tau_i)=\int_{0}^{\tau_i} d\tau'/a (\tau')$ be conformal time (where $d\Xi=d\tau/a(t)$), so that~\footnote{Technically, we should restrict our attention to $w>-1/3$ so that $\Xi$ is finite, but the relations presented below will still be valid if they are assumed to represent the integral evaluated at the upper limit alone and $w \neq -1/3$.} 
\begin{equation}
\int_{\tau_{I}}^{\tau_{\Lambda}} \frac{d\tau}{a_w} = \Xi (\tau_{\Lambda}) - \Xi (\tau_I).
\end{equation}
For any epoch lasting many Hubble times during which the equation of state is $p=w\rho$, with $w \neq -1/3$ (we will treat a possible epoch of curvature domination below), we have $(\rho (\tau_i) /\rho (\tau_j))=(a (\tau_i) /a (\tau_j))^{-3(1+w)}$ and $a (\tau_i)/a (\tau_j)=(\Xi (\tau_i)/\Xi (\tau_j))^{2/(3w+1)}$. Combining these gives
\be
{a(\tau_i)^2 \rho (\tau_i) \over a(\tau_j)^2 \rho (\tau_j)} = \left({\Xi (\tau_j) \over \Xi (\tau_i) }\right)^2.
\ee
Using the Friedmann Equation,
\be\label{eq-fe}
\Omega^{-1}-1={3\over 8\pi G \rho a^2},
\ee
we see that 
\begin{equation}
{\Xi (\tau_i) \over \Xi (\tau_j) } = \frac{\left( \Omega(\tau_i)^{-1}-1 \right)^{1/2}}{\left( \Omega(\tau_j)^{-1}-1 \right)^{1/2}},
\end{equation}
as long as $\Omega$ is not much less than one, which yields
\begin{equation}
\int_{\tau_{I}}^{\tau_{\Lambda}} \frac{d\tau}{a_w} = \Xi (\tau_I) \left[ \frac{\left( \Omega(\tau_{\Lambda})^{-1}-1 \right)^{1/2}}{\left( \Omega(\tau_I)^{-1}-1 \right)^{1/2}} -1 \right].
\end{equation}
Comparing $aH$ with $\Xi$ for a scale factor $a \propto \tau^{2 / 3(1+w)}$ we can express $ \Xi (\tau_I) $ as
\begin{equation}
\Xi (\tau_I) = \frac{2}{1+3w} \left(1 -  \Omega(\tau_{I}) \right)^{1/2},
\end{equation}
where the normalization is defined by the epoch immediately following inflation (via the value of $w$). This yields
\begin{eqnarray}
\int_{\tau_{I}}^{\tau_{\Lambda}} \frac{d\tau}{a_w} &=& \frac{2}{1+3w} \Omega (\tau_I)^{1/2} \\ 
&\times& \left[ \left( \Omega(\tau_{\Lambda})^{-1}-1 \right)^{1/2} -\left( \Omega(\tau_I)^{-1}-1 \right)^{1/2} \right] \nonumber.
\end{eqnarray}
From Eq.~\ref{eq-fe}, we have 
\be
\label{eq-omep}
\Omega (\tau_I)^{-1}-1 = {3 H_I^2\over 8\pi G\rho_I \sinh^2 H_I \tau_I} \approx \frac{e^{-2N_e}}{4},
\ee
where $N_e = H_I \tau_I$ is the number of inflationary efoldings. Assuming an appreciable number of efoldings, if $w<-1/3$, then $\Omega(\tau_{\Lambda}) >  \Omega(\tau_I)$ and the integral will be vanishingly small. If $w>-1/3$, then $\Omega(\tau_{\Lambda}) <  \Omega(\tau_I)$, and assuming an appreciable number of efolds, we obtain
\begin{equation}
\int_{\tau_{I}}^{\tau_{\Lambda}} \frac{d\tau}{a_w} \simeq \frac{2}{1+3w} \left( \Omega(\tau_{\Lambda})^{-1}-1 \right)^{1/2}.
\end{equation}
This expression is valid only while the curvature is not a significant component of the energy density ($\Omega$ is close to one), and we conclude that this contribution to ${\cal I}$ is necessarily small. However, if there is a period of curvature domination during some part of the cosmological evolution, say between $\tau_c > \tau_I$ and $\tau_{\Lambda}$, then we have $a \propto \tau$, which from the Friedmann equation yields (again, assuming an instantaneous transition between the various epochs)
\begin{equation}
\frac{\tau_{\Lambda}}{\tau_{c}} = \left[ \frac{(\Omega(\tau_{\Lambda})^{-1} -1)}{(\Omega(\tau_c)^{-1} -1)} \right]^{1/(1+3w)}
\end{equation}
where $w$ is defined by the component which was dominant before curvature. The contribution to the integral ${\cal I}$ is given by
\begin{equation}
\int_{\tau_c}^{\tau_{\Lambda}} \frac{d\tau}{a_{-1/3} (\tau)} = \frac{2(1-\Omega(\tau_c) )^{1/2} }{(1+w)(1+3w)} \log \left[\frac{(\Omega(\tau_{\Lambda})^{-1} -1)}{(\Omega(\tau_c)^{-1} -1)}  \right]
\end{equation}
which grows logarithmically for $w > -1/3$ as $\Omega(\tau_{\Lambda}) \rightarrow 0$, and remains small if $w < -1/3$ since $\Omega(\tau_{\Lambda}) \rightarrow 1$ in this case.
 
We now turn to the second integral in Eq.~\ref{eq-ii}, which after substituting with $a_{\Lambda}$ from Eq.~\ref{eq-epochs}, becomes
\begin{equation}
\int_{\tau_{\Lambda}}^{\tau_o} \frac{d\tau}{a_{\Lambda}} = \left(1- \Omega(\tau_{\Lambda}) \right)^{1/2} \left[ 1- e^{-H_{\Lambda} (\tau_o - \tau_{\Lambda} )}  \right].
\end{equation}
Allowing $\tau_o \rightarrow \infty$, the integral Eq.~\ref{eq-ii} becomes
\begin{eqnarray}
{\cal I} =&&  \frac{2}{1 +3w} \left( \Omega(\tau_{c})^{-1}-1 \right)^{1/2} \nonumber \\
&& + \frac{2(1-\Omega(\tau_c) )^{1/2} }{(1+w)(1+3w)} \log \left[\frac{(\Omega(\tau_{\Lambda})^{-1} -1)}{(\Omega(\tau_c)^{-1} -1)}  \right] \nonumber \\
&&   + \left(1- \Omega(\tau_{\Lambda}) \right)^{1/2}
\end{eqnarray}
Because $\Omega \rightarrow 1$ during the epoch of late-time vacuum energy, $\Omega(\tau_{\Lambda})$ is a minimum. It is this minimal value of $\Omega$ that determines the size of ${\cal I}$, which can be large only when there is a contribution from the term corresponding to a period of curvature domination before late-time vacuum energy domination. This has not happened in our universe, but we can determine whether it does in a more general model. Curvature domination occurs when 
\be
\rho \sim 3/8\pi G a^2,
\ee
and setting this to be the late-time vacuum energy gives 
\be
H^2_\Lambda \sim a^{-2} \approx \left({H_\Lambda\over H_{\rm eq}}\right)^{4/3}\left({H_{\rm eq}\over H_I}\right)a_I^{-2},
\ee
yielding
\be\label{eq-Hmin}
H_\Lambda \sim H_{\rm eq}^{-1/2}H_I^{3/2}e^{-3N_e},
\ee
where `eq' refers to the matter-radiation equality time.

Bubbles with vacuum energy larger than this will never see any significant enhancement of ${\cal I}$. We can use the value of $T_{\rm co}$, Eq.~\ref{Tcosimp}, to draw the conformal structure of a one-bubble spacetime more precisely: by following a null line from $T_{\rm co}$ into the bubble, we can determine the height of the ``hat" protruding above future infinity of the background de Sitter spacetime.  Note that because the late time vacuum energy enters in $T_{\rm co}$ only indirectly (by determining the minimal value of $\Omega$) a smaller vacuum energy does not necessarily guarantee a larger hat.

Returning to Eq.~\ref{Tcosimp}, we can now discuss the implications for observing late-time collisions. The relevant four-volume for bubble nucleation in the past light cone of an observer at rest with respect to the steady-state frame is (see Ref.~\cite{Garriga:2006hw} and~[AJS] for the relevant formulae)
\begin{eqnarray}
\label{eq-v4}
V_{4} &=& \frac{4 \pi}{3} H_{F}^{-4} \left[ \tan^2 T_{\rm co} +2 \log \left(1+ \tan T_{\rm co} \right)  \right] \nonumber \\ 
&=& \frac{4 \pi}{3 H_{F}^2 H_{I}^2} \left[ \tanh^2 \left( \frac{H_I \tau_I}{2}  \right) e^{2{\cal I}} \right.  \\ 
&& \left. + \frac{2 H_I^2}{H_F^2}  \log \left(1+ \frac{H_F}{H_I} \tanh \left( \frac{H_I \tau_I}{2}  \right) e^{{\cal I}} \right)  \right]  \nonumber
\end{eqnarray}
which reduces to the case presented in AJS when $H_F \gg H_I$ and ${\cal I} \rightarrow 0$, corresponding to a bubble filled with vacuum energy for all time. In {\em our} universe, we have ${\cal I} \simeq 0$ since current cosmological data favors a universe where $\Omega \simeq 1$, implying that $T_{\rm co} \simeq \arctan (H_{F}/H_I)$. To expect at least one collision in our past light cone, the nucleation probability must satisfy
\begin{equation}\label{eq-lambdabound}
\lambda H_F^{-4} > \frac{3}{4 \pi} \left( \frac{H_F^2}{H_I^2} + 2 \log \left[ 1+ \frac{H_F}{H_I} \right] \right)^{-1},
\end{equation}
which almost saturates the bound for eternal inflation unless $H_F$ is significantly larger than $H_I$; note that in this case collisions would enter the observer's lightcone during the bubble's inflation epoch and so is ``late time" only relative to $H_F^{-1}$.

Additional insight can be gleaned using the rate $dN/d\tau$ of such incoming bubbles. Using Eq.~\ref{eq-v4},
$$
{dN \over d\tau}  =  {\lambda dV_4 \over d\tau} = \lambda{dV_4 \over d{\cal I}}{d{\cal I}\over d\tau} \simeq 2 \left({4\pi\lambda\over 3H_F^4}\right)\left({H_F\over H_I}\right)^2e^{2{\cal I}}{d{\cal I}\over d\tau}, 
$$
assuming $H_I\tau_I \agt 1$ and $H_F \agt H_I$.  As noted above, $I \simeq 0$ unless $\Omega \ll 1$, and from its definition, $d{\cal I}/d\tau = 1/a$.  Thus
$$
{dN \over d\tau} \approx 2 \left({4\pi\lambda\over 3H_F^4}\right)\left({H_F\over H_I}\right)^2 a^{-1}.
$$
From this we see that of order $\lambda H_F^{-4}(H_F/H_I)^2$ bubbles enter the lightcone during the first few efolds of inflation; thereafter the rate declines exponentially.  At late times, from the Hubble equation, $1/a \sim H(\Omega^{-1}-1)^{1/2}$ prior to $\Lambda$-domination, so even with the nucleation rate near saturation, bubbles (or their causal effects) will collide with a late-time observer more than once per Hubble time only to the degree that $(1-\Omega)^{1/4}$ exceeds $H_I/H_F$.

In terms of size on the sky, the expected angular scale of a collision is governed by the probability distributions derived in AJS, and is determined by $T_{\rm co}$. For $T_{\rm co} \sim \pi / 4$, the maximum of the distribution (which is rather wide) is located at $\psi_{m} \simeq .75$. The maximum falls to an angular scale of one degree when $T_{\rm co} \simeq \pi / 2 - x$ with $x = .05$. 

\end{appendix}

\bibliography{thinwallcollisions}

\end{document}